\RequirePackage{fix-cm}
\documentclass[smallextended]{svjour3}       
\smartqed  
\usepackage{graphicx,natbib,amssymb}
\journalname{Celestial Mechanics \& Dynamical Astronomy}

\begin{document}

\title{Explicit evolution relations with orbital elements for eccentric, inclined, elliptic and hyperbolic restricted few-body problems
}

\titlerunning{Relations for the General Restricted Problem}        

\author{Dimitri Veras}


\institute{Dimitri Veras \at
              Department of Physics, University of Warwick, Gibbet Hill Road, Coventry CV4 7AL \\
              Tel.: +44 (024) 765 23965\\
              Fax:  +44 (024) 761 50897\\
              \email{d.veras@warwick.ac.uk}           
}
\date{Received: 24 November 2013 / Revised: 14 January 2014 / Accepted: 16 January 2014 /}

\maketitle

\begin{abstract}
Planetary, stellar and galactic physics often rely on the general restricted gravitational $N$-body problem to model the motion of a small-mass object under the influence of much more massive objects.  Here, I formulate the general restricted problem entirely and specifically in terms of the commonly-used orbital elements of semimajor axis, eccentricity, inclination, longitude of ascending node, argument of pericentre, and true anomaly, without any assumptions about their magnitudes.  I derive the equations of motion in the general, unaveraged case, as well as specific cases, with respect to both a bodycentric and barycentric origin.  I then reduce the equations to three-body systems, and present compact singly- and doubly-averaged expressions which can be readily applied to systems of interest.  This method recovers classic Lidov-Kozai and Laplace-Lagrange theory in the test particle limit to any order, but with fewer assumptions, and reveals a complete analytic solution for the averaged planetary pericentre precession in coplanar circular circumbinary systems to at least the first three nonzero orders in semimajor axis ratio.  Finally, I show how the unaveraged equations may be used to express resonant angle evolution in an explicit manner that is not subject to expansions of eccentricity and inclination about small nor any other values.
\end{abstract}

\section{Overview}

The movement of an infinitesimal mass in a region dominated by massive bodies
has important implications for designing spacecraft missions \citep{gometal2001},
preparing for near-Earth interlopers (Shoemaker 1995; de la Fuente
Marcos \& de la Fuente Marcos 2013), and 
understanding the behaviour of planetary, stellar and galactic 
systems (Binney \& Tremaine 1987; Murray \& Dermott 1999).  
Applications are far-reaching 
\citep[e.g. to black holes,][]{schnittman2010}.  This resulting motion is 
related, but not strictly equivalent, to the motion found in the
{\it restricted problem}.

\subsection{Context}

The seminal work of \cite{szebehely1967} claims to be ``the first book devoted
to the theory of orbits in the restricted problem''.  His historical perspective
highlights the inherent assumptions which accompany the term {\it restricted}
and have since been reinforced by later celestial mechanics texts
(pg. 253 of \citealt{danby1992}, pg. 63 of \citealt{murder1999}).  These assumptions
are i) the system contains three bodies, two of which are massive, ii) the
two massive bodies share a mutual circular orbit, and iii) all three bodies have coplanar
orbits.  Other texts have begun to explicitly use the terms {\it circular} or
{\it planar} to qualify the otherwise broad terminology 
(pg. 196 of \citealt{morbidelli2002}, pg. 118 of \citealt{roy2005}, 
pg. 115 of \citealt{valkar2006})

Standard treatments of this famous but quite specific case follow a similar pattern
of deriving the Cartesian equations of motion by introducing a rotating coordinate 
system and defining a potential from which zero-velocity surfaces, the Jacobi constant
(or integral of motion), the Tisserand parameter and the five Lagrangian equilibrium 
points may be obtained.  Although the resulting equations of motion provide insight 
into several concepts, such as the Hill sphere, they do not immediately shed light on
some basic orbit characteristics such as how the pericentre of the zero-mass body 
changes with time.

\subsection{Objective}

In this paper, I derive the equations of motion for the {\it general} restricted
problem in terms of solely the semimajor axis, $a$, eccentricity, $e$, inclination, $i$, 
longitude of ascending node, $\Omega$, argument of pericentre,
$\omega$, and true anomaly, $f$ of the zero-mass body and of all of the massive bodies.  
I will also use the mean motion 
$n$ as a convenient auxiliary parameter than can be expressed solely
in terms of $a$ and the masses.  The word {\it general} refers to the
removal of all the aforementioned assumptions; the systems here may host
an arbitrary number of massive bodies on arbitrary but known orbits.
By no means, however, is this case the most general type of restricted problem
(see Chapter 1.9 of \citealt{szebehely1967} for other extensions).

Importantly, I present the unaveraged equations as well as the averaged
equations; the latter case for the unrestricted 3-body problem has been scrutinized 
in depth-recently, largely since the initial discovery and confirmation of 
extrasolar planets \citep{wolfra1992,wolszczan1994}. The general restricted 
equations I present here are not confined to small mass ratios (as is characteristic of
the related work of Henri Poincar\'{e}), small perturbative forces, nor any type of
expansion about a limiting orbital element value.  The resulting relations
may be potentially useful tools which can be applied to a problem of interest.
Such problems need not contain a test particle; as long as the smallest mass
is much smaller than the other masses, the equations will describe the
motion to a good approximation.

\subsection{Benefits of orbital element approach}

Small-body, planetary and stellar dynamicists often rely on the set of elements
$(a,e,i,\Omega,\omega)$ to obtain an intuitive feel for the osculating motion.
The location of an object along its orbit can be gleaned from $f$, or alternatively
the mean anomaly, mean longitude or true longitude.  All these elements directly
demonstrate, for example, how close or how far an object may extend from a
massive body, and are easily amenable to limiting cases.  For example, classic 
Lidov-Kozai theory, which assumes the presence of a test particle, is based on 
the interplay between $e$ and $i$.  Also, one reason
why the Tisserand parameter is so useful is because it relates $a$, $e$ and $i$
to one another.

Observational data is another major motivation for using orbital elements.  The
majority of extrasolar planets have been discovered by Doppler radial velocity
spectroscopy, which yields an observable from which $e$, $\omega$ and $f$ could
be measured with a fit to the data.  Further, the three major exoplanet databases
(see the Extrasolar Planets Encyclopedia at http://exoplanet.eu/,
the Exoplanet Data Explorer at http://exoplanets.org/ and
the NASA Exoplanet Archive at http://exoplanetarchive.ipac.caltech.edu/) all
report data in terms of orbital elements.  Finally, for purposes of direct
integration of a known stellar or planetary system, avoiding scaled Cartesian 
coordinates removes the need to convert both the input and output.

\begin{figure}
\centerline{
  \includegraphics[width=1.00\textwidth,height=15cm]{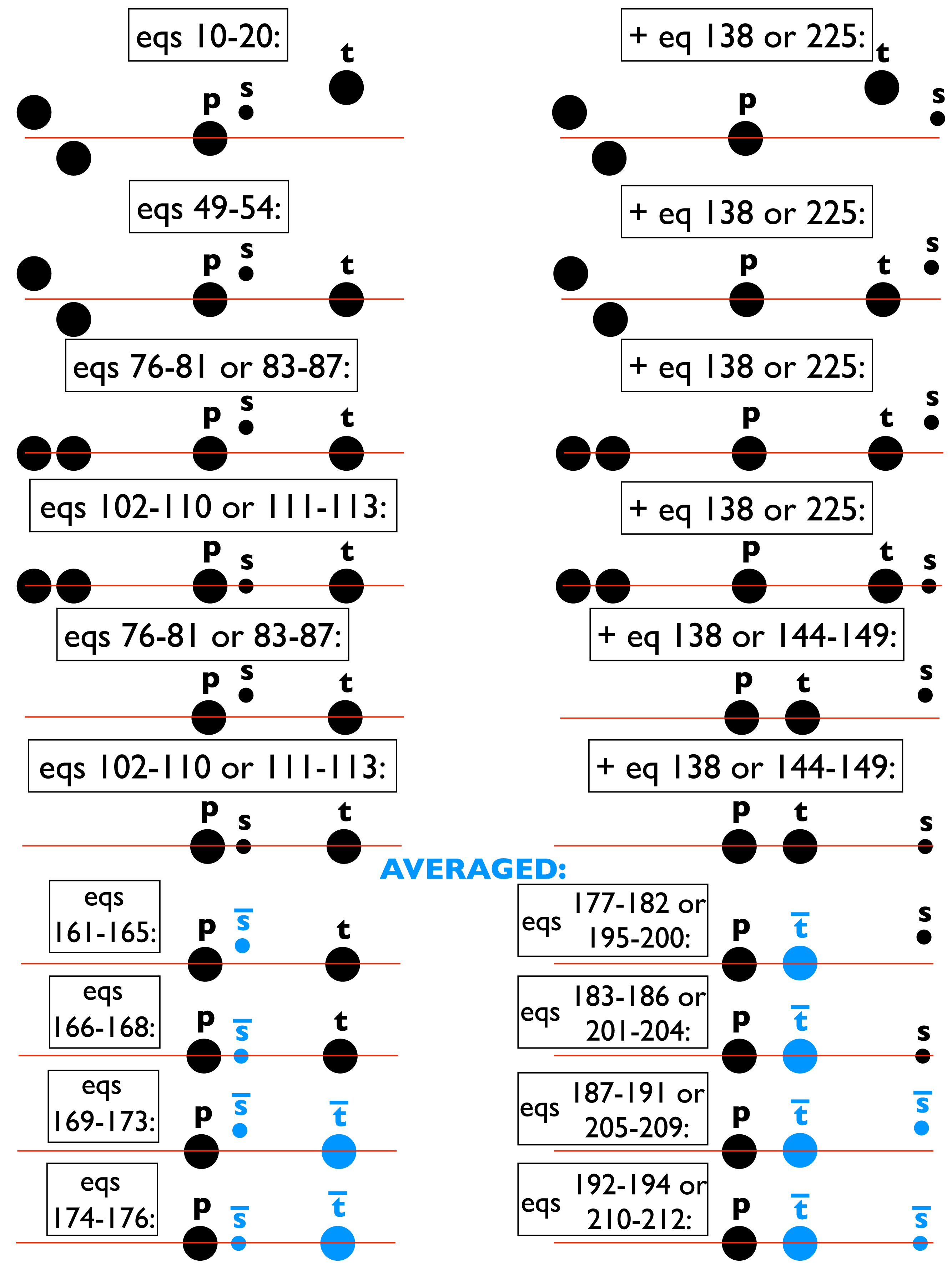}
}
\caption{Representative cartoons of different restricted $N$-body problems considered in this paper, along with the corresponding equation numbers describing the equations of motion.  Here ``p'', ``s'' and ``t'' refer to the primary, secondary and tertiary; the secondary is always massless and all other bodies are always massive.  Although a fourth and fifth body are present in the top 8 configurations, these bodies are merely a proxy for an arbitrary number of bodies.  Each red line is the orbital plane of the primary; other bodies placed on that line share elliptic or hyperbolic coplanar orbits with the primary.  The left column refers to setups where the orbital elements are measured with respect to the primary (typically when the other massive bodies are exterior to the secondary), and the right column where the orbital elements are measured with respect to the barycentre of a given number of massive bodies (typically when the secondary's orbit is exterior to more than one massive body).  A blue body containing an overbar indicates that body's orbit is averaged over its true anomaly.  Averaged tertiary orbits are assumed to be elliptical.
}
\label{TOC}   
\end{figure}

\subsection{How to use this paper}

The reader can use the equations in this paper i) for direct integration to solve
for the motion of the zero-mass body\footnote{No integration is necessary for equation (\ref{solveexact}).}, 
ii) to obtain physical intuition for what 
orbital properties are the most significant catalysts of orbital variation, iii) to 
treat a wide variety of restricted problems in a consistent analytical
framework, and iv) to derive existing theories in an alternate manner.
The only key assumptions made throughout the paper is that the object I classify
as the secondary contains no mass and an osculating elliptical orbit, and that the orbits 
of all other bodies are known functions of time.

The reader should first identify the number of bodies in their restricted problem, assumptions
about their orbits, and the reference point from which to measure orbital elements.  Then
scanning Fig. \ref{TOC} will help identify the appropriate
setup.  Each red line in the figure refers to the orbital plane of the massive
primary ``p''.  The left column features setups where the orbital elements of the 
massless secondary ``s'' is measured with respect to the primary; in the right
column the secondary's orbital elements are measured with respect to the barycentre
of more than one of the massive bodies.  The top eight configurations generally refer
to the $N$-body problem (not specifically the 5-body problem), and the bottom eight 
configurations all showcase averaged elliptical orbits.

\subsection{Outline of paper}

That figure provides specific equation numbers, but here I describe the content
of the various sections.  First, I set up the problem in Section \ref{setup} 
before describing the derivation technique in Section \ref{dervt}.  The next
three sections (\ref{genine}-\ref{fullcop}) present the equations of motion
for, first, an arbitrary number of bodies on arbitrary orbits, then when one reference
plane of one or more of the massive bodies is fixed, and finally for the assumption
that all bodies have forever coplanar orbits.  These equations all assume
that the orbital elements are measured with respect to the primary.
Section \ref{threebod} briefly touches on what modifications to the equations
can be made when only three bodies are in the system.

The paper then transitions, and evaluates how the equations would be 
transformed if orbital elements were measured with respect to some
barycentric reference frame.  Section \ref{barythree} presents the
three-body case, and Appendix A presents the general case.  Section
\ref{barythree} contains both the necessary scaling form and the
explicit equations themselves.

Up until that point, all equations considered will have been unaveraged,
and contain the true anomalies of all of the bodies in the system.
Sections \ref{avgsec}-\ref{circum} consider averaged cases for the
three-body problem. I consider every type of averaging for an
internal (Section \ref{scsec}) and external (Sections \ref{sfsec}-\ref{circum})
secondary.  Section \ref{circum} considers the relevant and analytically
tractable case of a primary-tertiary pair on a circular orbit.
A brief exposition on resonances follows in Section \ref{ress}, and Section 
\ref{summ} summarizes this work.

\section{Setup} \label{setup}

Consider a system that contains $N \ge 3$ gravitationally interacting point 
masses $m_j$, where gravity is the only acting force, and $j = 3...N$.  
Assume the position of the secondary 
with respect to the primary is denoted by $\vec{r}=(x,y,z)$
and the position of all other bodies with respect to 
the primary by $\vec{r}_j = (x_j,y_j,z_j)$.  The massive primary
($m_1 \equiv m_{\rm p}$) and massless secondary ($m_2 \equiv m_{\rm s} = 0$)
are assumed to be initially bound to one another.  The tertiary
mass is denoted by $m_3 \equiv m_{\rm t}$.  In effect, the equations
of motion can be applied for a relatively small but nonzero $m_s$ to
an excellent approximation. An example of one configuration is
a Solar-type star (primary), an asteroid (secondary), a terrestrial planet (tertiary) 
and a Jovian planet (quaternary), where the motions of the tertiary
and quaternary about the primary are known.

\section{Derivation technique} \label{dervt}

The general restricted system contains no known constants of the motion.  
Neither energy nor angular momentum is conserved.  The Jacobi constant
and the Tisserand parameter do not apply, except in a specific
case.  Without these tools to help derive the equations, I instead turn to perturbation
theory, where the perturbation may be arbitrary large.

Lagrange's planetary equations are useful here because they are derived 
without approximation \citep[e.g.][]{brocle1961}.  Other derivations, such
as for evolution equations described by given radial, tangential and
normal components of a perturbative force                                          
(Burns 1976 and pgs. 54-57 of 
Murray \& Dermott 1999) are not used
because they assume that the perturbed force is small.  Lagrange's planetary
equations traditionally contain a truncated disturbing function, but need
not.  The equations can instead be expressed as equation (22) of \cite{efroimsky2005} 
and equation (16) of \cite{gurfil2007}, for an arbitrary perturbative 
acceleration and in terms of precomputed matrices of Poisson Brackets
and partial derivatives of positions with respect to orbital elements
\footnote{The derivation of
Lagrange's planetary equations contains a previously missed degree
of freedom \citep{efrgol2003,efrgol2004}, which, although not exploited
here, may be applied in future studies to obtain new insight into the motion.}.
The relevant equations can be found in \cite{vereva2013} and are not repeated here.

The form of the perturbative acceleration is the key to application of the method.
The acceleration must be a function of the position and velocity of the
secondary only, and must be a simple enough function of the positions and velocities
to be analytically tractable.  Denote this acceleration as $\Delta$.  Then

\begin{equation}
\underbrace{
\frac{d^2\vec{r}}{dt^2} 
= 
- \frac{G \left(m_p + m_s \right) \vec{r}}{r^3}
}_{\rm classic \ 2-body \ problem}
\
\
+
\underbrace{
\Delta
}_{\rm perturbation}
\label{eqmotion}
\end{equation}

\noindent{}where the arbitrarily large perturbative accelerations on the secondary orbit
are $\Delta = \sum_{j=3}^{N} \left(  \Delta_{j,A} + \Delta_{j,B}\right)$
where

\begin{equation}
\Delta_{j,A} = - \frac{G m_{j} \vec{r}_{j}}{r_{j}^3}
\label{deltaja}
\end{equation}

\noindent{and}

\begin{equation}
\Delta_{j,B} = \frac{G m_{j} \left(\vec{r}_{j} - \vec{r}\right)}
{\left|\vec{r}_{j} - \vec{r}\right|^3}
.
\label{deltajb}
\end{equation}

\noindent{}Although $\Delta$ contains $\vec{r}_j$ terms, they are, crucially,
independent of the secondary's position and velocity because the secondary has
zero mass. Further, I find that the functional
dependence on $(x,y,z)$ is not complex enough to prevent the method from succeeding.
The time evolution of the secondary's orbital
elements is additive so that they can be decomposed into separate terms attributable
to both $\Delta_{j,A}$ and $\Delta_{j,B}$.  I obtain

\begin{eqnarray}
\frac{da}{dt} &=& \sum_{j=3}^{N} \left[\left(\frac{da}{dt}\right)_{j,A} + \left(\frac{da}{dt}\right)_{j,B} \right]
,
\label{firstdadt}
\\
\frac{de}{dt} &=& \sum_{j=3}^{N} \left[\left(\frac{de}{dt}\right)_{j,A} + \left(\frac{de}{dt}\right)_{j,B} \right]
,
\\
\frac{di}{dt} &=& \sum_{j=3}^{N} \left[\left(\frac{di}{dt}\right)_{j,A} + \left(\frac{di}{dt}\right)_{j,B} \right]
,
\\
\frac{d\Omega}{dt} &=& \sum_{j=3}^{N} \left[\left(\frac{d\Omega}{dt}\right)_{j,A} + \left(\frac{d\Omega}{dt}\right)_{j,B} \right]
,
\\
\frac{d\omega}{dt} &=& \sum_{j=3}^{N} \left[\left(\frac{d\omega}{dt}\right)_{j,A} + \left(\frac{d\omega}{dt}\right)_{j,B} \right]
,
\\
\frac{df}{dt} &=& \left(\frac{df}{dt}\right)_{\rm unperturbed \ 2-body} +
\sum_{j=3}^{N} \left[\left(\frac{df}{dt}\right)_{j,A} + \left(\frac{df}{dt}\right)_{j,B} \right]
.
\end{eqnarray}

\noindent{}The unperturbed two-body term describes the orbital evolution of the classic two-body problem. In order to derive the desired equations from $\Delta$, I follow the same algebraic procedure described in \cite{vereva2013}.  Now I begin presenting the results.

\section{General equations in the inertial frame} \label{genine}

In the general restricted $N$-body problem, the equations of motion for the massless secondary's orbit are

\begin{eqnarray}
\left(\frac{da}{dt}\right)_{j,A} &=&
  \frac{2Gm_{j}}{n\sqrt{1-e^2}r_{j}^3} 
\big[
x_{j} \left(C_1 \cos{i} \sin{\Omega} + C_2 \cos{\Omega} \right) 
\nonumber
\\
&+&
y_{j} \left(-C_1 \cos{i} \cos{\Omega} + C_2 \sin{\Omega} \right)
-
z_{j} \left(C_1 \sin{i}\right)
\big]
,
\label{dadtGENA}
\\
\left(\frac{de}{dt}\right)_{j,A} &=&
  \frac{Gm_{j}\sqrt{1-e^2}}{2an\left(1 + e\cos{f}\right)r_{j}^3} 
\big[
x_{j} \left(C_6 \cos{i} \sin{\Omega} + C_5 \cos{\Omega} \right) 
\nonumber
\\
&+&
y_{j} \left(-C_6 \cos{i} \cos{\Omega} + C_5 \sin{\Omega} \right)
-
z_{j} \left(C_6 \sin{i}\right)
\big]
,
\label{dedtGENA}
\\
\left(\frac{di}{dt}\right)_{j,A} &=&
  -\frac{Gm_{j}\sqrt{1-e^2}}{an\left(1 + e\cos{f}\right)r_{j}^3} 
\cos{\left(f + \omega\right)}
\big[
x_{j} \left(\sin{i} \sin{\Omega} \right) 
\nonumber
\\
&-&
y_{j} \left(\sin{i} \cos{\Omega} \right) +
z_{j} \left(\cos{i}\right)
\big]
,
\label{didtGENA}
\\
\left(\frac{d\Omega}{dt}\right)_{j,A} &=&
  -\frac{Gm_j\sqrt{1-e^2}}{an\left(1 + e\cos{f}\right)r_{j}^3} 
\sin{\left(f + \omega\right)}
\big[
x_{j} \left(\sin{\Omega} \right) 
\nonumber
\\
&-&
y_{j} \left(\cos{\Omega} \right) +
z_{j} \left(\cot{i}\right)
\big]
,
\label{dOdtGENA}
\\
\left(\frac{d\omega}{dt}\right)_{j,A} &=&
  \frac{Gm_{j}\sqrt{1-e^2}}{2aen\left(1 + e\cos{f}\right)r_{j}^3} 
\big[
x_{j} \left(-C_8 \cos{i} \sin{\Omega} + C_7 \cos{\Omega} \right) 
\nonumber
\\
&+&
y_{j} \left(C_8 \cos{i} \cos{\Omega} + C_7 \sin{\Omega} \right) 
\nonumber
\\
&+& 
z_{j} \left(C_9 \sin{i} + 2e \sin{\left(f + \omega\right)} \cos{i}\cot{i} \right)
\big]
\label{dodtGENA}
\end{eqnarray}

\noindent{and}

\begin{eqnarray}
\left(\frac{da}{dt}\right)_{j,B} &=&
  \frac{2Gm_{j}}{n\sqrt{1-e^2}r_{j,B}^3} 
\big[
x_{j,B} \left(-C_1 \cos{i} \sin{\Omega} - C_2 \cos{\Omega} \right) 
\nonumber
\\
&+&
y_{j,B} \left(C_1 \cos{i} \cos{\Omega} - C_2 \sin{\Omega} \right) +
z_{j,B} \left(C_1 \sin{i}\right)
\big]
,
\label{dadtGENB}
\\
\left(\frac{de}{dt}\right)_{j,B} &=&
  \frac{Gm_{j}\sqrt{1-e^2}}{2an\left(1 + e\cos{f}\right)r_{j,B}^3} 
\big[ - 2a \sin{f} \left(1 - e^2\right)
\nonumber
\\
&+&
x_j \left(-C_6 \cos{i} \sin{\Omega} - C_5 \cos{\Omega} \right) 
\nonumber
\\
&+&
y_{j} \left(C_6 \cos{i} \cos{\Omega} - C_5 \sin{\Omega} \right) 
+
z_{j} \left(C_6 \sin{i}\right)
\big]
,
\label{dedtGENB}
\\
\left(\frac{di}{dt}\right)_{j,B} &=&
  \frac{Gm_{j}\sqrt{1-e^2}}{an\left(1 + e\cos{f}\right)r_{j,B}^3} 
\cos{\left(f + \omega\right)}
\nonumber
\\
&\times&
\left[
x_{j} \left(\sin{i} \sin{\Omega} \right) -
y_{j} \left(\sin{i} \cos{\Omega} \right) +
z_{j} \left(\cos{i}\right)
\right]
,
\label{didtGENB}
\\
\left(\frac{d\Omega}{dt}\right)_{j,B} &=&
  \frac{Gm_j\sqrt{1-e^2}}{an\left(1 + e\cos{f}\right)r_{j,B}^3} 
\sin{\left(f + \omega\right)}
\nonumber
\\
&\times&
\left[
x_{j} \left(\sin{\Omega} \right) -
y_{j} \left(\cos{\Omega} \right) +
z_{j} \left(\cot{i}\right)
\right]
,
\label{dOdtGENB}
\\
\left(\frac{d\omega}{dt}\right)_{j,B} &=&
  \frac{Gm_j\sqrt{1-e^2}}{2aen\left(1 + e\cos{f}\right)r_{j,B}^3} 
\nonumber
\\
&\times&
\bigg[
-2e \cos{i} \sin{\left(f + \omega\right)}
\left[  
x_{j} \left(\sin{\Omega} \right) -
y_{j} \left(\cos{\Omega} \right) +
z_{j} \left(\cot{i}\right)
\right]
\nonumber
\\
&+&
x_{j,B} \left(C_9 \cos{i} \sin{\Omega} - C_7 \cos{\Omega} \right) 
\nonumber
\\
&+&
y_{j,B} \left(-C_9 \cos{i} \cos{\Omega} - C_7 \sin{\Omega} \right) +
z_{j,B} \left(-C_9 \sin{i}\right)
\bigg]
\label{dodtGENB}
\end{eqnarray}

\noindent{with} 

\begin{equation}
\frac{df}{dt} = \frac{n \left(1 + e \cos{f}\right)^2}{\left(1 - e^2\right)^{3/2}}
                - \frac{d\omega}{dt} - \cos{i} \frac{d\Omega}{dt}
.
\label{dfdt}
\end{equation}

\noindent{}The auxiliary set of $C$ variables depend only
on the orbital parameters of the primary-secondary orbit
and can be expressed as

\begin{eqnarray}
C_1 &\equiv& e \cos{\omega} + \cos{\left(f+\omega\right)}
,
\label{C1}
\\
C_2 &\equiv& e \sin{\omega} + \sin{\left(f+\omega\right)}
,
\label{C2}
\\
C_3 &\equiv& \cos{i} \sin{\Omega} \sin{\left(f+\omega\right)} 
    - \cos{\Omega} \cos{\left(f + \omega \right)}
,
\label{C3}
\\
C_4 &\equiv& \cos{i} \cos{\Omega} \sin{\left(f+\omega\right)} 
    + \sin{\Omega} \cos{\left(f + \omega \right)}
,
\label{C4}
\\
C_5 &\equiv& \left(3 + 4e \cos{f} + \cos{2f} \right) \sin{\omega}
        + 2 \left( e + \cos{f} \right) \cos{\omega} \sin{f}
,
\label{C5}
\\
C_6 &\equiv& \left(3 + 4e \cos{f} + \cos{2f} \right) \cos{\omega}
        - 2 \left( e + \cos{f} \right) \sin{\omega} \sin{f}
,
\label{C6}
\\
C_7 &\equiv& \left(3 + 2e \cos{f} - \cos{2f} \right) \cos{\omega}
        + \sin{\omega} \sin{2f}
,
\label{C7}
\\
C_8 &\equiv& \left(3  - \cos{2f} \right) \sin{\omega}
        - 2 \left(e + \cos{f} \right) \cos{\omega} \sin{f} 
,
\label{C8}
\\
C_9 &\equiv& \left(3 + 2e \cos{f} - \cos{2f} \right) \sin{\omega}
        - \cos{\omega} \sin{2f}
.
\label{C9}
\end{eqnarray}

\noindent{The} Cartesian components of the position vectors of all of the
massive bodies in orbital elements are

\begin{eqnarray}
x_{j} &=& r_{j} \left[\cos{\Omega_{j}} \cos{\left(f_{j} + \omega_{j}\right)} 
                    - \sin{\Omega_{j}} \sin{\left(f_{j} + \omega_{j}\right)} \cos{i_{j}} \right]
,
\label{xjGEN}
\\
y_{j} &=& r_{j} \left[\sin{\Omega_{j}} \cos{\left(f_{j} + \omega_{j}\right)} 
                    + \cos{\Omega_{j}} \sin{\left(f_{j} + \omega_{j}\right)} \cos{i_{j}} \right]
,
\label{yjGEN}
\\
z_{j} &=& r_{j} \left[\sin{\left(f_{j} + \omega_{j}\right)} \sin{i_{j}} \right] 
\label{zjGEN}
\end{eqnarray}

\noindent{with} 

\begin{equation}
r_{j} = \frac{p_{j}}{1 + e_{j} \cos{f_{j}}}
\label{rjGEN}
\end{equation}

\noindent{}where for elliptical and hyperbolic orbits,
$p_{j} = a_{j} \left(1 - e_{j}^2\right)$,
and $p_{j} = a_{j} \left(e_{j}^2 - 1\right)$, respectively.
For a parabolic tertiary orbit, $p_{j}$ equals twice the 
pericentric distance. The difference between an elliptic
and hyperbolic restricted problem resides simply in the
definition of $r_{j}$ in equation (\ref{rjGEN}).
Also, $\vec{r}_{j,B} \equiv \left(x_{j,B}, y_{j,B}, z_{j,B} \right)$
with

\begin{eqnarray}
x_{j,B} &=& x_{j} + \frac{aC_3 \left(1 - e^2\right)}{1 + e\cos{f}}
= x_{j} + r C_3
,
\label{xjB}
\\
y_{j,B} &=& y_{j} - \frac{aC_4 \left(1 - e^2\right)}{1 + e\cos{f}}
= y_{j} - r C_4
,
\label{yjB}
\\
z_{j,B} &=& z_{j} - \frac{a \left(1 - e^2\right) \sin{i} \sin{\left(f+\omega\right)}}{1 + e\cos{f}}
= z_{j} - r \sin{i} \sin{\left(f +\omega\right)}
.
\label{zjB}
\end{eqnarray}

\noindent{}Now the equations of motion have been expressed entirely in terms
of orbital elements.
I use the definitions of the $C$ variables in order to maintain consistency with 
\cite{vereva2013}.  The form of equations (\ref{xjGEN})-(\ref{zjGEN}) makes no assumptions
about the boundedness of the orbit for the tertiary, as the true 
anomaly $f_{j}$ can be defined for all orbit types
as the angle between the pericentre and the tertiary's location.
Usually, for parabolic and hyperbolic orbits, the reference direction is
coplanar with the orbit and coincides with the line between
the pericentre and the primary.  Hence, in those contexts, the
angles $\omega_{j}$ and $\varpi_{j}$ are rarely used.

\section{General equations in the rotated frame} \label{rotframe}

\subsection{A fixed primary-tertiary reference plane}

I can simplify the equations of motion by tilting 
the reference frame so that it coincides
with the plane of the two-body
orbit between the primary and one of the $m_j$, $j \ge 3$ 
bodies.  Here I use the primary-tertiary orbital plane as 
the reference plane, with an arbitrary but fixed
reference direction within that plane to measure 
the orbital angles.

This transformation, however, comes at a cost. In order
for the equations to be most useful, I must assume that the primary-tertiary
orbit plane remains fixed in space and does not precess due
to the influence of the bodies denoted by $m_j$, $j \ge 4$.
In reality, the plane will precess and the reference direction will
change by some nonzero amount because the $j \ge 4$
bodies are not massless.  However, the precession is often negligible
in several realistic cases, such as the ecliptic of the Solar System,
and four and five-body problems which are hierarchical
in mass (for example, a restricted three-body problem contained within a
restricted four-body problem)\footnote{The quality of the approximation may
be estimated by considering the precession rate of the primary-tertiary orbital
plane in the solution of the full three-body problem with the primary, tertiary 
and the most massive body $m_j$, $j \ge 4$.}.  Therefore, although the 
equations in this section for systems with $N \ge 4$ bodies are 
technically inexact, they may prove useful.

Consequently, the orbital parameters of all other bodies are now 
measured with respect to this (assumed-fixed) orbital plane
and reference direction.
If viewed face-on and if the primary and tertiary orbit each other, 
then the orbital motion of the primary
and tertiary can be in one of two directions.
As viewed from the north poles of those objects,
assume that they orbit in a counterclockwise fashion.
Then I can set $i_{\rm t} = 0^{\circ}$ and 
$\varpi_{\rm t} = \omega_{\rm t} + \Omega_{\rm t}$,
where $\varpi$ represents the longitude of 
pericentre\footnote{Alternatively, for clockwise motion I can 
set $i_{\rm t} = 180^{\circ}$ and define an {\it obverse of pericentre}
as in \cite{vereva2013}.}.  This action allows me to
eliminate $i_{\rm t}$, $\omega_{\rm t}$ and $\Omega_{\rm t}$
from the equations such that now

\begin{eqnarray}
x_{\rm t} &=& r_{\rm t} \cos{\left(f_{\rm t} + \varpi_{\rm t}\right)}
,
\label{xtROT}
\\
y_{\rm t} &=& r_{\rm t} \sin{\left(f_{\rm t} + \varpi_{\rm t}\right)}
,
\label{ytROT}
\\
z_{\rm t} &=& 0
.
\label{ztROT}
\end{eqnarray}

\noindent{and}

\begin{eqnarray}
r_{3,B} \equiv r_{{\rm t},B} 
&=&
r_{\rm t} {\left\lbrace  
1 - \kappa_{\rm t} \left( \frac{r}{r_{\rm t}} \right)
+ \left(\frac{r}{r_{\rm t}}\right)^2
\right\rbrace}^{1/2}
\label{rtBfar}
\\
&=&
r {\left\lbrace  
1 - \kappa_{\rm t} \left( \frac{r_{\rm t}}{r} \right)
+ \left(\frac{r_{\rm t}}{r}\right)^2
\right\rbrace}^{1/2}
\label{rtBnear}
.
\end{eqnarray}

\noindent{}I write $r_{t,B}$ in both these forms to foreshadow future expansions
of this variable when averaging the equations of motion.  In either case,

\begin{eqnarray}
\kappa_{\rm t} &=& 
2
\left[  
\cos{\left(f + \omega\right)}
\cos{\chi_{\rm t}}
-
\cos{i}
\sin{\left(f + \omega\right)}
\sin{\chi_{\rm t}}
\right]
\nonumber
\\
&=&
2 \left(D_{{\rm t},2} \cos{f} - D_{{\rm t},1} \sin{f} \right)
\label{kappafar}
\\
&=&
2 \left(D_{{\rm t},3} \cos{f_{\rm t}} + D_{{\rm t},4} \sin{f_{\rm t}} \right)
\label{kappanear}
\end{eqnarray}

\noindent{with}

\begin{eqnarray}
\chi_{\rm t} &=&  \Omega - f_{\rm t} - \varpi_{\rm t}
,
\label{chit}
\\
D_{{\rm t},1} &=& 
    \sin{\omega} \cos{\chi_{\rm t}}
    +
    \cos{\omega} \cos{i} \sin{\chi_{\rm t}}
,
\label{D1}
\\
D_{{\rm t},2} &=& 
    \cos{\omega} \cos{\chi_{\rm t}}
    -
    \sin{\omega} \cos{i} \sin{\chi_{\rm t}}
,
\label{D2}
\\
D_{{\rm t},3} &=& \cos{\left(f+ \omega \right)} \cos{\left(\Omega - \varpi_{\rm t}\right)}
        -
        \cos{i}\sin{\left(f+ \omega \right)} \sin{\left(\Omega - \varpi_{\rm t}\right)}
,
\label{D3}
\\
D_{{\rm t},4} &=& \cos{\left(f+ \omega \right)} \sin{\left(\Omega - \varpi_{\rm t}\right)}
        +
        \cos{i}\sin{\left(f+ \omega \right)} \cos{\left(\Omega - \varpi_{\rm t}\right)}
.
\label{D4}
\end{eqnarray}

Now I can re-express the equations of motion as

\begin{eqnarray}
\frac{da}{dt} &=&
  \frac{2Gm_{\rm t}r_{\rm t}}{n\sqrt{1-e^2}} 
\left( \frac{S_{({\rm t},A,a)}}{r_{{\rm t}}^3} + \frac{S_{({\rm t},B,a)}}{r_{{\rm t},B}^3} \right) 
\nonumber
\\
&+&
\sum_{j=4}^{N} \left[\left(\frac{da}{dt}\right)_{j,A} + \left(\frac{da}{dt}\right)_{j,B} \right]
,
\label{dadtROT1}
\\
\frac{de}{dt} &=&
  \frac{Gm_{\rm t}\sqrt{1-e^2}r_{\rm t}}{2an\left(1 + e\cos{f}\right)} 
\left( \frac{S_{({\rm t},A,e)}}{r_{{\rm t}}^3} + \frac{S_{({\rm t},B,e)}}{r_{{\rm t},B}^3} \right) 
\nonumber
\\
&+&
\sum_{j=4}^{N} \left[\left(\frac{de}{dt}\right)_{j,A} + \left(\frac{de}{dt}\right)_{j,B} \right]
,
\label{dedtROT1}
\\
\frac{di}{dt} &=&
  \frac{Gm_{\rm t}\sqrt{1-e^2}r_{\rm t}}{an\left(1 + e\cos{f}\right)} 
\left( \frac{S_{({\rm t},A,i)}}{r_{{\rm t}}^3} + \frac{S_{({\rm t},B,i)}}{r_{{\rm t},B}^3} \right) 
\nonumber
\\
&+&
\sum_{j=4}^{N} \left[\left(\frac{di}{dt}\right)_{j,A} + \left(\frac{di}{dt}\right)_{j,B} \right]
,
\label{didtROT1}
\\
\frac{d\Omega}{dt} &=&
  \frac{Gm_{\rm t}\sqrt{1-e^2}r_{\rm t}}{an\left(1 + e\cos{f}\right)}
 \left( \frac{S_{({\rm t},A,\Omega)}}{r_{{\rm t}}^3} + \frac{S_{({\rm t},B,\Omega)}}{r_{{\rm t},B}^3} \right) 
\nonumber
\\
&+&
\sum_{j=4}^{N} \left[\left(\frac{d\Omega}{dt}\right)_{j,A} + \left(\frac{d\Omega}{dt}\right)_{j,B} \right]
,
\label{dOdtROT1}
\\
\frac{d\omega}{dt} &=&
  \frac{Gm_{\rm t}\sqrt{1-e^2}r_{\rm t}}{2aen\left(1 + e\cos{f}\right)}
  \left( \frac{S_{({\rm t},A,\omega)}}{r_{{\rm t}}^3} + \frac{S_{({\rm t},B,\omega)}}{r_{{\rm t},B}^3} \right) 
\nonumber
\\
&+&
\sum_{j=4}^{N} \left[\left(\frac{d\omega}{dt}\right)_{j,A} + \left(\frac{d\omega}{dt}\right)_{j,B} \right]
,
\label{dodtROT1}
\\
\frac{df}{dt} &=& \frac{n \left(1 + e \cos{f}\right)^2}{\left(1 - e^2\right)^{3/2}}
                - \frac{d\omega}{dt} - \cos{i} \frac{d\Omega}{dt}
.
\label{dfdtROT1}
\end{eqnarray}

\noindent{}where all contributions from the $j \ge 4$ bodies are given in equations (\ref{dadtGENA})-(\ref{dodtGENB}).  The auxiliary $S$ variables, defined below, are particularly useful ways to characterize the relative contributions from both the $A$ and $B$ terms.  I express the $S$ variables alternatively in terms of $f$ and $f_j$ to facilitate later study of configurations when the secondary is both interior and exterior to the tertiary.  I find

\begin{eqnarray}
S_{(j,A,a)} &=& 
 e D_{j,1}
 +D_{j,1}\cos{f}
 +D_{j,2}\sin{f} 
\label{SjAa}
\\
&=& 
 D_{j,5}\cos{f_j}
 -D_{j,6}\sin{f_j}
,
\\
S_{(j,A,e)} &=& 
   3 D_{j,1}
  + 2eD_{j,2}\sin{f}
  + 4eD_{j,1}\cos{f}
\nonumber
\\
  &&+ D_{j,2}\sin{2f}
  + D_{j,1}\cos{2f}
\\
&=& 
 D_{j,7}\cos{f_{j}}
 -D_{j,8}\sin{f_{j}}
,
\\
S_{(j,A,i)} &=& \sin{i} \sin{\chi_j}
\big\lbrace 
\sin{f} \sin{\omega}
-
\cos{f} \cos{\omega}
\big\rbrace
\\
&=& \sin{i} \cos{\left(f+\omega\right)}
\big\lbrace 
\cos{\left(\Omega - \varpi_{j}\right)}\sin{f_{j}}
-
\sin{\left(\Omega - \varpi_{j}\right)}\cos{f_{j}} 
\big\rbrace
,
\\
S_{(j,A,\Omega)} &=& - \sin{\chi_j}
\big\lbrace 
\sin{\omega}\cos{f}
+
\cos{\omega}\sin{f}
\big\rbrace
\\
&=& \sin{\left(f+\omega\right)}
\big\lbrace 
\cos{\left(\Omega - \varpi_{j}\right)}\sin{f_{j}}
-
\sin{\left(\Omega - \varpi_{j}\right)}\cos{f_{j}}
\big\rbrace
,
\\
S_{(j,A,\omega)} &=& 
   3 D_{j,2}
  + \left(2e\cos{i}\cos{\omega}\sin{\chi_j}\right)\sin{f}
\nonumber
\\
 &&+ \left(2e\cos{\omega}\cos{\chi_j}\right)\cos{f}
  - D_{j,2}\cos{2f}
  + D_{j,1}\sin{2f}
\\
&=&  
D_{j,10}\sin{f_{j}}
-
D_{j,9}\cos{f_{j}}
\end{eqnarray}

\noindent{and}

\begin{eqnarray}
S_{(j,B,a)}  &=& - S_{(j,A,a)} - \left(\frac{r}{r_{j}}\right) e \sin{f}
,
\label{SjBa}
\\
S_{(j,B,e)}  &=& - S_{(j,A,e)} - \left(\frac{r}{r_{j}}\right) \left(2 \sin{f} + e \sin{2f} \right)
,
\label{SjBe}
\\
S_{(j,B,i)}  &=& -S_{(j,A,i)}
,
\label{SjBi}
\\
S_{(j,B,\Omega)}  &=& -S_{(j,A,\Omega)}
,
\label{SjBO}
\\
S_{(j,B,\omega)}  &=& - S_{(j,A,\omega)} + \left(\frac{r}{r_{j}}\right) \left(2 \cos{f} + 2e \cos^2{f} \right)
,
\label{SjBo}
\end{eqnarray}

\noindent{with}

\begin{eqnarray}
D_{{\rm j},5} &=& 
    C_1 \cos{i} \sin{\left(\Omega - \varpi_j\right)}
    +
    C_2 \cos{\left(\Omega - \varpi_j\right)}
,
\label{D5}
\\
D_{{\rm j},6} &=& 
    C_1 \cos{i} \cos{\left(\Omega - \varpi_j\right)}
    -
    C_2 \sin{\left(\Omega - \varpi_j\right)}
,
\label{D6}
\\
D_{{\rm j},7} &=& 
    C_6 \cos{i} \sin{\left(\Omega - \varpi_j\right)}
    +
    C_5 \cos{\left(\Omega - \varpi_j\right)}
,
\label{D7}
\\
D_{{\rm j},8} &=& 
    C_6 \cos{i} \cos{\left(\Omega - \varpi_j\right)}
    -
    C_5 \sin{\left(\Omega - \varpi_j\right)}
,
\label{D8}
\\
D_{{\rm j},9} &=& 
    C_8 \cos{i} \sin{\left(\Omega - \varpi_j\right)}
    -
    C_7 \cos{\left(\Omega - \varpi_j\right)}
,
\label{D9}
\\
D_{{\rm j},10} &=& 
    C_8 \cos{i} \cos{\left(\Omega - \varpi_j\right)}
    +
    C_7 \sin{\left(\Omega - \varpi_j\right)}
.
\label{D10}
\end{eqnarray}

The lack of an extra term on the RHS of both equations (\ref{SjBi}) and (\ref{SjBO})
explains why the inclination and longitude of ascending node are more simply expressed
analytically than the eccentricity and longitude of pericentre.  The variation of the
inclination and longitude of ascending node have a symmetry about both perturbative
terms (equations \ref{deltaja}-\ref{deltajb}) that is lacking from the eccentricity and argument of pericentre.

\subsection{When all massive bodies are coplanar}

If, however, all
bodies in the system except perhaps the secondary are coplanar and 
the pericentres of the coplanar bodies (from elliptical,
parabolic or hyperbolic orbits) are all measured with respect to the
same reference direction, then the equations of motion reduce to

\begin{eqnarray}
\frac{da}{dt} &=&
\sum_{j=3}^{N}
  \frac{2Gm_jr_j}{n\sqrt{1-e^2}} 
\left( \frac{S_{(j,A,a)}}{r_{j}^3} + \frac{S_{(j,B,a)}}{r_{j,B}^3} \right) 
,
\label{dadtROT2}
\\
\frac{de}{dt} &=&
\sum_{j=3}^{N}
  \frac{Gm_j\sqrt{1-e^2}r_{j}}{2an\left(1 + e\cos{f}\right)} 
\left( \frac{S_{(j,A,e)}}{r_{j}^3} + \frac{S_{(j,B,e)}}{r_{j,B}^3} \right) 
,
\label{dedtROT2}
\\
\frac{di}{dt} &=&
\sum_{j=3}^{N}
  \frac{Gm_j\sqrt{1-e^2}r_{j}}{an\left(1 + e\cos{f}\right)} 
\left( \frac{S_{(j,A,i)}}{r_{j}^3} + \frac{S_{(j,B,i)}}{r_{j,B}^3} \right) 
,
\label{didtROT2}
\\
\frac{d\Omega}{dt} &=&
\sum_{j=3}^{N}
  \frac{Gm_j\sqrt{1-e^2}r_{j}}{an\left(1 + e\cos{f}\right)}
 \left( \frac{S_{(j,A,\Omega)}}{r_{j}^3} + \frac{S_{(j,B,\Omega)}}{r_{j,B}^3} \right) 
,
\label{dOdtROT2}
\\
\frac{d\omega}{dt} &=&
\sum_{j=3}^{N}
  \frac{Gm_j\sqrt{1-e^2}r_{j}}{2aen\left(1 + e\cos{f}\right)}
  \left( \frac{S_{(j,A,\omega)}}{r_{j}^3} + \frac{S_{(j,B,\omega})}{r_{j,B}^3} \right) 
,
\label{dodtROT2}
\\
\frac{df}{dt} &=& \frac{n \left(1 + e \cos{f}\right)^2}{\left(1 - e^2\right)^{3/2}}
                - \frac{d\omega}{dt} - \cos{i} \frac{d\Omega}{dt}
.
\label{dfdtROT2}
\end{eqnarray}

\noindent{}The remainder of the paper utilizes equations (\ref{dadtROT2})-(\ref{dfdtROT2}) or reductions of these equations.

\subsubsection{Preparations for averaging}

The equations may be recast in a form that will be useful when one (later) wishes
to obtain averaged effects.  Typically, averaging can be performed only
if the quantity $r_{3,B}$ is expanded about a dimensionless ratio 
$\alpha_{\rm sc} \equiv r/r_{\rm t}$ or 
$\alpha_{\rm sf} \equiv r_{\rm t}/r$ in a power series, where the subscripts indicate
if the secondary is {\it close} or {\it far} from the primary.  Hence, the following 
identity is useful.

\begin{equation}
\left(1 - \kappa_{\rm t} \alpha_{\rm sc} + \alpha_{\rm sc}^2 \right)^{-3/2}
=
\sum_{u = 0}^{\infty} 
     \left(
     \begin{array}{c}
     \frac{1}{2} + u \\
     \frac{1}{2}
     \end{array}
     \right)
     \alpha_{\rm sc}^u
     \left[\kappa_{\rm t} - \alpha_{\rm sc}\right]^u
\label{series}
\end{equation}

\noindent{}Consequently, the equations of motion may be expressed as a power series in $\alpha_{\rm sc}$ or $\alpha_{\rm sf}$.  In the former case, by using equations (\ref{SjBa})-(\ref{SjBo}), I obtain

\begin{eqnarray}
\frac{da}{dt} &=&
\sum_{j=3}^{N}
  \frac{-2Gm_{j}}{n\sqrt{1-e^2}r_{j}^2} 
\bigg[
S_{(j,A,a)} 
     \sum_{u = 1}^{\infty} 
     \left(
     \begin{array}{c}
     \frac{1}{2} + u \\
     \frac{1}{2}
     \end{array}
     \right)
     \alpha_{\rm sc}^u
     \left(\kappa_j - \alpha_{\rm sc}\right)^u
\nonumber
\\
&+&
e \sin{f}
     \sum_{u = 0}^{\infty} 
     \left(
     \begin{array}{c}
     \frac{1}{2} + u \\
     \frac{1}{2}
     \end{array}
     \right)
     \alpha_{\rm sc}^{u+1}
     \left(\kappa_j - \alpha_{\rm sc}\right)^u
\bigg]
,
\label{avgcopa}
\\
\frac{de}{dt} &=&
\sum_{j=3}^{N}
  \frac{-Gm_{j}\sqrt{1-e^2}}{2an\left(1 + e\cos{f}\right)r_{j}^2} 
\bigg[
S_{(j,A,e)} 
     \sum_{u = 1}^{\infty} 
     \left(
     \begin{array}{c}
     \frac{1}{2} + u \\
     \frac{1}{2}
     \end{array}
     \right)
     \alpha_{\rm sc}^u
     \left(\kappa_j - \alpha_{\rm sc}\right)^u
\nonumber
\\
&+& 
\left(2\sin{f} + e \sin{2f} \right)
     \sum_{u = 0}^{\infty} 
     \left(
     \begin{array}{c}
     \frac{1}{2} + u \\
     \frac{1}{2}
     \end{array}
     \right)
     \alpha_{\rm sc}^{u+1}
     \left(\kappa_j - \alpha_{\rm sc}\right)^u
\bigg]
,
\\
\frac{di}{dt} &=&
\sum_{j=3}^{N}
  \frac{-Gm_{j}\sqrt{1-e^2}}{an\left(1 + e\cos{f}\right)r_{j}^2} 
S_{(j,A,i)} 
     \sum_{u = 1}^{\infty} 
     \left(
     \begin{array}{c}
     \frac{1}{2} + u \\
     \frac{1}{2}
     \end{array}
     \right)
     \alpha_{\rm sc}^u
     \left(\kappa_j - \alpha_{\rm sc}\right)^u
,
\\
\frac{d\Omega}{dt} &=&
\sum_{j=3}^{N}
  \frac{-Gm_j\sqrt{1-e^2}}{an\left(1 + e\cos{f}\right)r_{j}^2}
S_{(j,A,\Omega)} 
     \sum_{u = 1}^{\infty} 
     \left(
     \begin{array}{c}
     \frac{1}{2} + u \\
     \frac{1}{2}
     \end{array}
     \right)
     \alpha_{\rm sc}^u
     \left(\kappa_j - \alpha_{\rm sc}\right)^u
,
\\
\frac{d\omega}{dt} &=&
\sum_{j=3}^{N}
  \frac{Gm_j\sqrt{1-e^2}}{2aen\left(1 + e\cos{f}\right)r_{j}^2}
\bigg[
-S_{(j,A,\omega)} 
     \sum_{u = 1}^{\infty} 
     \left(
     \begin{array}{c}
     \frac{1}{2} + u \\
     \frac{1}{2}
     \end{array}
     \right)
     \alpha_{\rm sc}^u
     \left(\kappa_j - \alpha_{\rm sc}\right)^u
\nonumber
\\
&+& 
2\left(\cos{f} + e \cos^2{f} \right)
     \sum_{u = 0}^{\infty} 
     \left(
     \begin{array}{c}
     \frac{1}{2} + u \\
     \frac{1}{2}
     \end{array}
     \right)
     \alpha_{\rm sc}^{u+1}
     \left(\kappa_j - \alpha_{\rm sc}\right)^u
\bigg]
.
\label{avgcopom}
\end{eqnarray}

\noindent{}Note the difference in the starting summation indices.  The power series in $\alpha_{\rm sf}$ is not particularly useful here because typically when the secondary is exterior to the tertiary, one no longer wishes to measure the secondary's orbit elements with respect to the primary.

\section{Full coplanarity of all bodies} \label{fullcop}

If I now impose coplanarity on the secondary as well, then the equations
are greatly simplified.  I set $i = 0^{\circ}$ and $\varpi = \omega + \Omega$.
Consequently, the auxiliary {\it planar} $C$ variables, denoted by $C^{\rm P}$, take on the same form except with the
substitution $\omega \rightarrow \varpi$.  The two exceptions are
$C_{3}^{\rm P} = -\cos{\left(f + \varpi\right)}$ and 
$C_{4}^{\rm P} = \sin{\left(f + \varpi\right)}$.  Hence,

\begin{eqnarray}
x_{j,B} &=&  r_{j} \cos{\left(f_{j} + \varpi_{j}\right)}
-
            r \cos{\left(f + \varpi\right)}
,
\\
y_{j,B} &=&  r_{j} \sin{\left(f_{j} + \varpi_{j}\right)}
-
            r \sin{\left(f + \varpi\right)}
,
\\
z_{j,B} &=&   0
,
\end{eqnarray}

\noindent{with}

\begin{eqnarray}
x_{j} &=& r_{j} \cos{\left(f_{j} + \varpi_{j}\right)}
,
\\
y_{j} &=& r_{j} \sin{\left(f_{j} + \varpi_{j}\right)}
,
\\
z_{j} &=& 0
,
\end{eqnarray}

\noindent{}and

\begin{eqnarray}
\chi_{j}^{\rm P} &=& \varpi - f_j - \varpi_j
,
\\
\kappa_{j}^{\rm P} &=& 2 \cos{\left(f + \chi_{j}^{\rm P}\right)}
\\
&=& 2 \left(D_{j,2}^{\rm P} \cos{f} - D_{j,1}^{\rm P} \sin{f}\right) 
\\
&=& 2 \left(D_{j,3}^{\rm P}\cos{f_j}  + D_{j,4}^{\rm P} \sin{f_j} \right) 
\end{eqnarray}

\noindent{such} that

\begin{eqnarray}
D_{j,1}^{\rm P} &=&  \sin{\chi_{j}^{\rm P}}
,
\\
D_{j,2}^{\rm P} &=&  \cos{\chi_{j}^{\rm P}}
,
\\
D_{j,3}^{\rm P} &=&  \cos{\left(f + \varpi - \varpi_j\right)}
,
\\
D_{j,4}^{\rm P} &=&  \sin{\left(f + \varpi - \varpi_j\right)} 
.
\end{eqnarray}

\noindent{}The fully coplanar equations of motion, which are denoted with a superscript P, become

\begin{eqnarray}
\left(\frac{da}{dt}\right)_{j,A}^{\rm P} &=&
  \frac{2Gm_j}{n\sqrt{1-e^2}r_{j}^3} 
\left[
x_{j} C_{2}^{\rm P} - y_{j} C_{1}^{\rm P}
\right]
,
\label{dadtPLANGENA}
\\
\left(\frac{de}{dt}\right)_{j,A}^{\rm P} &=&
  \frac{Gm_{j}\sqrt{1-e^2}}{2an\left(1 + e\cos{f}\right)r_{j}^3} 
\left[
x_{j} C_{5}^{\rm P} - y_{j} C_{6}^{\rm P}   
\right]
,
\label{dedtPLANGENE}
\\
\left(\frac{di}{dt}\right)_{j,A}^{\rm P} &=& 0
,
\label{didtPLANGENI}
\\
\left(\frac{d\varpi}{dt}\right)_{j,A}^{\rm P} &=&
  \frac{Gm_{j}\sqrt{1-e^2}}{2aen\left(1 + e\cos{f}\right)r_{j}^3} 
\left[
x_{j} C_{7}^{\rm P} +
y_{j} C_{9}^{\rm P}
\right]
,
\label{dcdtPLANGENA}
\end{eqnarray}

\noindent{}and

\begin{eqnarray}
\left(\frac{da}{dt}\right)_{j,B}^{\rm P} &=&
  \frac{2Gm_{j}}{n\sqrt{1-e^2}r_{j,B}^3} 
\nonumber
\\
&\times&
\left[
\frac{-ae\left(1 - e^2\right)\sin{f}}{1 + e \cos{f}}  -x_{j} C_{2}^{\rm P} + y_{j} C_{1}^{\rm P}
\right]
,
\\
\left(\frac{de}{dt}\right)_{j,B}^{\rm P} &=&
  \frac{Gm_{j}\sqrt{1-e^2}}{2an\left(1 + e\cos{f}\right)r_{j,B}^3} 
\nonumber
\\
&\times&
\left[
-2a\left(1 - e^2\right) \sin{f} - x_{j} C_{5}^{\rm P} + y_{j} C_{6}^{\rm P}   
\right]
,
\\
\left(\frac{di}{dt}\right)_{j,B}^{\rm P} &=& 0
,
\\
\left(\frac{d\varpi}{dt}\right)_{j,B}^{\rm P} &=&
  \frac{Gm_{j}\sqrt{1-e^2}}{2aen\left(1 + e\cos{f}\right)r_{j,B}^3} 
\nonumber
\\
&\times&
\left[
2a\left(1 - e^2 \right) \cos{f}
-x_{j} C_{7}^{\rm P} -
y_{j} C_{9}^{\rm P}
\right]
.
\end{eqnarray}

\noindent{}with

\begin{equation}
\left(\frac{df}{dt}\right)^{\rm P} = \frac{n \left(1 + e \cos{f}\right)^2}{\left(1 - e^2\right)^{3/2}}
- \left(\frac{d\varpi}{dt}\right)_{j,A}^{\rm P} - \left(\frac{d\varpi}{dt}\right)_{j,B}^{\rm P}
.
\label{dfdtplanar}
\end{equation}

\noindent{}I find here, as in \cite{vereva2013}, that $(d\Omega/dt)^{\rm P} \ne 0$
even though the orbits remain coplanar.  This feature has no physical
consequence, but is likely mathematically important to include if 
$\omega$ was kept in the equations instead of $\varpi$. 

\subsection{Preparations for averaging}

I eliminate the $C^{\rm P}$ terms and reexpress the terms
in square brackets for the fully coplanar equations of motion as

\begin{eqnarray}
\left(\frac{da}{dt}\right)^{\rm P} &=&
\sum_{j=3}^{N}
  \frac{-2Gm_{j}}{n\sqrt{1-e^2}r_{j}^2} 
\bigg[
S_{(j,A,a)}^{\rm P} 
     \sum_{u = 1}^{\infty} 
     \left(
     \begin{array}{c}
     \frac{1}{2} + u \\
     \frac{1}{2}
     \end{array}
     \right)
     \alpha_{\rm sc}^u
     \left(\kappa_{j}^{\rm P} - \alpha_{\rm sc}\right)^u
\nonumber
\\
&+&
e \sin{f}
     \sum_{u = 0}^{\infty} 
     \left(
     \begin{array}{c}
     \frac{1}{2} + u \\
     \frac{1}{2}
     \end{array}
     \right)
     \alpha_{\rm sc}^{u+1}
     \left(\kappa_{j}^{\rm P} - \alpha_{\rm sc}\right)^u
\bigg]
,
\label{altcopa}
\\
\left(\frac{de}{dt}\right)^{\rm P} &=&
\sum_{j=3}^{N}
  \frac{-Gm_{j}\sqrt{1-e^2}}{2an\left(1 + e\cos{f}\right)r_{j}^2} 
\bigg[
S_{(j,A,e)}^{\rm P} 
     \sum_{u = 1}^{\infty} 
     \left(
     \begin{array}{c}
     \frac{1}{2} + u \\
     \frac{1}{2}
     \end{array}
     \right)
     \alpha_{\rm sc}^u
     \left(\kappa_{j}^{\rm P} - \alpha_{\rm sc}\right)^u
\nonumber
\\
&+&
\left(2\sin{f} + e \sin{2f} \right)
     \sum_{u = 0}^{\infty} 
     \left(
     \begin{array}{c}
     \frac{1}{2} + u \\
     \frac{1}{2}
     \end{array}
     \right)
     \alpha_{\rm sc}^{u+1}
     \left(\kappa_{j}^{\rm P} - \alpha_{\rm sc}\right)^u
\bigg]
,
\\
\left(\frac{d\varpi}{dt}\right)^{\rm P} &=&
\sum_{j=3}^{N}
  \frac{Gm_j\sqrt{1-e^2}}{2aen\left(1 + e\cos{f}\right)r_{j}^2}
\bigg[
-S_{(j,A,\varpi)}^{\rm P} 
     \sum_{u = 1}^{\infty} 
     \left(
     \begin{array}{c}
     \frac{1}{2} + u \\
     \frac{1}{2}
     \end{array}
     \right)
     \alpha_{\rm sc}^u
     \left(\kappa_{j}^{\rm P} - \alpha_{\rm sc}\right)^u
\nonumber
\\
&+&
2\left(\cos{f} + e \cos^2{f} \right)
     \sum_{u = 0}^{\infty} 
     \left(
     \begin{array}{c}
     \frac{1}{2} + u \\
     \frac{1}{2}
     \end{array}
     \right)
     \alpha_{\rm sc}^{u+1}
     \left(\kappa_{j}^{\rm P} - \alpha_{\rm sc}\right)^u
\bigg]
,
\label{altcopom}
\end{eqnarray}

\noindent{where}

\begin{eqnarray}
S_{(j,A,a)}^{\rm P}
&=&
e D_{j,1}^{\rm P}
+ D_{j,1}^{\rm P} \cos{f}
+ D_{j,2}^{\rm P} \sin{f}
\\
&=&
D_{j,5}^{\rm P}
\cos{f_j}
-
D_{j,6}^{\rm P}
\sin{f_j}
,
\\
S_{(j,A,e)}^{\rm P}
&=&
  3 D_{j,1}^{\rm P}
+4 e D_{j,1}^{\rm P} \cos{f}
+    D_{j,1}^{\rm P} \cos{2f}
\nonumber
\\
&&+ 2 e D_{j,2}^{\rm P} \sin{f}
+    D_{j,2}^{\rm P} \sin{2f}
\\
&=&
D_{j,7}^{\rm P}
\cos{f_j}
-
D_{j,8}^{\rm P}
\sin{f_j}
,
\\
S_{(j,A,\varpi)}^{\rm P}
&=&
  3 D_{j,2}^{\rm P}
+ 2e D_{j,2}^{\rm P} \cos{f}
-    D_{j,2}^{\rm P} \cos{2f}
+    D_{j,1}^{\rm P} \sin{2f}
\\
&=&
D_{j,12}
\cos{f_j}
-
D_{j,11}
\sin{f_j}
,
\end{eqnarray}

\noindent{}and

\begin{eqnarray}
S_{(j,B,a)}^{\rm P} &=& 
-S_{(j,A,a)}^{\rm P}
- \left(\frac{r}{r_{j}}\right) e \sin{f}
,
\\
S_{(j,B,e)}^{\rm P} &=& 
-S_{(j,A,e)}^{\rm P}
- \left(\frac{r}{r_{\rm t}}\right) \left(2\sin{f} + e \sin{2f} \right)
,
\\
S_{(j,B,\varpi)}^{\rm P} &=& 
-S_{(j,A,\varpi)}^{\rm P}
+ \left(\frac{r}{r_{\rm t}}\right) \left(2\cos{f} + 2e \cos^2{f} \right)
,
\end{eqnarray}

\noindent{with} 

\begin{eqnarray}
D_{j,5}^{\rm P} &=&  e \sin{\left(\varpi - \varpi_{j}\right)} + \sin{\left(f+\varpi-\varpi_j\right)}
,
\\
D_{j,6}^{\rm P} &=&  e \cos{\left(\varpi - \varpi_{j}\right)} + \cos{\left(f+\varpi-\varpi_j\right)}
,
\\
D_{j,7}^{\rm P} &=& 
3 \sin{\left(\varpi - \varpi_{j}\right)} 
+
3 e \sin{\left(f + \varpi - \varpi_{j}\right)}
\nonumber
\\
&&+
\sin{\left(2f + \varpi - \varpi_{j}\right)}
-
e \sin{\left(f - \varpi + \varpi_{j}\right)}
,
\\
D_{j,8}^{\rm P} &=& 
3 \cos{\left(\varpi - \varpi_{j}\right)} 
+
3 e \cos{\left(f + \varpi - \varpi_{j}\right)}
\nonumber
\\
&&+
\cos{\left(2f + \varpi - \varpi_{j}\right)}
+
e \cos{\left(f - \varpi + \varpi_{j}\right)}
,
\\
D_{j,11} &=& 
\sin{\left(2f + \varpi - \varpi_j \right)}
-
\left(3 + 2 e \cos{f}\right)
\sin{\left(\varpi - \varpi_j\right)} 
,
\\
D_{j,12} &=& 
-\cos{\left(2f + \varpi - \varpi_j \right)}
+
\left(3 + 2 e \cos{f}\right)
\cos{\left(\varpi - \varpi_j\right)}
.
\end{eqnarray}

\section{Three bodies only} \label{threebod}

Here I briefly place three-body systems in context of the equations of motion already presented, before describing these systems in greater detail in later sections.  First, I note that the Tisserand parameter does not apply in the averaged systems which I describe later.

\subsection{An inclined and circular tertiary}

The equations of motion for this system are equations (\ref{dadtROT2})-(\ref{dfdtROT2}),
or, alternatively, (\ref{avgcopa})-(\ref{avgcopom}) plus equation (\ref{dfdtROT2}).
If I assume that the tertiary is on a circular (bound) orbit,
regardless of the orientation of the secondary orbit,
then I can both simplify the equations ($e_{\rm t} = 0$) and introduce
an additional constraint on the system through the
Tisserand parameter, $T$.  Then
%
%
the Tisserand parameter is conserved such that

\begin{eqnarray}
T &=& \frac{a_{\rm t}}{a}
+ 
2 \sqrt{\frac{ a \left(1 - e^2\right)}{a_{\rm t}}  }
\cos{i}
\nonumber
\\
&=&
\frac{1}{\alpha_{\rm sc}}
\left(\frac{1 - e^2}{1 + e \cos{f}} \right)
+
2\cos{i}\sqrt{\alpha_{\rm sc} \left(1 + e \cos{f}\right)} 
\label{Tiss}
\end{eqnarray}

\noindent{}which can help eliminate $a$, 
$e$, $i$ or $f$ from the equations of motion.  Further, 
because $T$ may be expressed in terms of $\alpha_{\rm sc}$,
averaging may be facilitated.

\subsection{A coplanar and circular tertiary} \label{3bod}

Here the equations of motion are equivalent to the
fully coplanar equations (equations \ref{dadtPLANGENA}-\ref{dfdtplanar}, or 
\ref{altcopa}-\ref{altcopom} plus \ref{dfdtplanar}) except that I
may also use the Tisserand parameter (equation \ref{Tiss})
with $i = 0$ to eliminate one variable.  

In this system, $r_{\rm t} = a_{\rm t}$, and equations (\ref{rtBfar})-(\ref{rtBnear}) hold,
except with 

\begin{eqnarray}
\kappa_{\rm t} &=& 2 \cos{\left(f + f_j + \varpi\right)}
\\
&=& 2 \left[ \cos{\left(f_j + \varpi\right)} \cos{f}
           - \sin{\left(f_j + \varpi\right)} \sin{f}  \right]
,
\\
&=& 2 \left[ \cos{\left(f + \varpi\right)} \cos{f_j}
           - \sin{\left(f + \varpi\right)} \sin{f_j}  \right]
.
\end{eqnarray}



\section{Shifting the reference origin} \label{barythree}

Until now, I have assumed that the secondary's orbital elements are 
measured with respect to the centre of the primary.  This assumption
generally yields useful elements when all other massive bodies in the system
are further away from the primary than the secondary \footnote{An exception
might be a Trojan asteroid of a planet.}.  

However, if the secondary orbits a cluster of massive bodies, then 
measuring the orbital elements with respect to just one of those
bodies, like the primary, will yield unhelpful relations.  The high reflex motion of 
the primary will make the secondary appear to oscillate between
elliptic and hyperbolic orbits, in many cases improperly giving 
the impression that the system is unstable.  The more natural
way to trace the secondary's motion is to compute the orbital elements
with respect to the centre of mass of the cluster.

Deriving these elements requires me to express the equation of motion
of the secondary and the centre of mass of the other bodies in the
same form as in equation (\ref{eqmotion}).  Subsequently, I could perform
the same perturbation analysis as in Sections \ref{genine}-\ref{threebod}.
However, there is a shortcut that enables me to derive the 
new equations more elegantly.

\subsection{General 3-body equations of motion}

To proceed, I restrict the derivation to three bodies here, and
describe the general $N$-body case in Appendix A.  Figure \ref{tridia}
shows the primary (``p''), secondary (``s'') and tertiary (``t''),
where $m_s = 0$, $m_p \ne 0$ and $m_t \ne 0$.  I need to convert 
equation (\ref{eqmotion}) into a similar equation for $\vec{s}$.
I have

\begin{equation}
\vec{s} = -\vec{s}_{\rm p} - \vec{r} = -\vec{s}_{\rm t} + \vec{r}_{\rm t} - \vec{r}
                               = -\left(\frac{m_{\rm p}}{m_{\rm p} +m_{\rm t}} \right)\vec{r}_{\rm t} + \vec{r}_{\rm t} - \vec{r}
\end{equation}

\noindent{}so that insertion into equation (\ref{eqmotion}) yields

\begin{eqnarray}
&-&\frac{d^2\vec{s}}{dt^2} +
 \left(\frac{m_t}{m_{\rm p} + m_{\rm t}} \right) \frac{d^2\vec{r}_{\rm t}}{dt^2}
+ G m_p \left[ 
              \frac{-\vec{s} + \left(\frac{m_{\rm t}}{m_{\rm p}+m_{\rm t}}\right) \vec{r}_{\rm t}}
             {\left| -\vec{s} + \left(\frac{m_{\rm t}}{m_{\rm p}+m_{\rm t}}\right) \vec{r}_{\rm t}  \right|^3}
        \right]
\nonumber
\\
&=& 
 G m_t \left[ 
              \frac{\vec{r}_{\rm t} + \vec{s} - \left(\frac{m_t}{m_{\rm p}+m_{\rm t}}\right) \vec{r}_{\rm t}}
             {\left|\vec{r}_{\rm t} + \vec{s} - \left(\frac{m_t}{m_{\rm p}+m_{\rm t}}\right) \vec{r}_{\rm t}  \right|^3}
       \right]
-
G m_t \frac{\vec{r}_{\rm t}}{r_{\rm t}^3}
.
\label{inter}
\end{eqnarray}

\noindent{}In order to eliminate $d^2\vec{r}_{\rm t}/dt^2$ from equation (\ref{inter}), I use the properties
of the two body problem (as the secondary has no mass) to write

\begin{equation}
\frac{d^2\vec{r}_{\rm t}}{dt^2} + G \left(m_p + m_{\rm t}\right) \frac{\vec{r}_{\rm t}}{r_{\rm t}^3} = 0
\end{equation}

\noindent{}so that by adding and subtracting the same term below, I finally obtain the desired form

\begin{eqnarray}
\frac{d^2\vec{s}}{dt^2} 
&=&
-
\frac{G\left(m_{\rm p} + m_{\rm t}\right)\vec{s}}{s^3}
+
G m_p \left[ 
              \frac{\left(\frac{m_{\rm t}}{m_{\rm p}+m_{\rm t}}\right) \vec{r}_{\rm t} - \vec{s}}
             {\left| \left(\frac{m_{\rm t}}{m_{\rm p}+m_{\rm t}}\right) \vec{r}_{\rm t} -\vec{s} \right|^3}
        \right]
\\
&-&
G m_t \left[ 
              \frac{\left(\frac{m_{\rm p}}{m_{\rm p}+m_{\rm t}}\right) \vec{r}_{\rm t} + \vec{s}}
             {\left| \left(\frac{m_{\rm p}}{m_{\rm p}+m_{\rm t}}\right) \vec{r}_{\rm t} + \vec{s} \right|^3}
        \right]
+
\frac{G \left( m_{\rm p} + m_{\rm t} \right)\vec{s}}{s^3}
,
\label{bary2body}
\end{eqnarray}

\noindent{}which can be compared to equation (\ref{eqmotion}).

This form demonstrates that the perturbation $\Delta'$ to the two-body problem in the barycentric reference frame is composed of three terms, all variations of $\Delta_{j,B}$ (equation \ref{deltajb}).  I denote the orbital elements in this reference frame with primes, and let $\beta$ and $\beta'$ represent placeholders for any of $(a,e,i,\Omega,\omega,f)$ and $(a',e',i',\Omega',\omega',f')$ respectively.  Then the barycentric equations of motion in orbital elements are given fully by 

\begin{eqnarray}
\frac{d\beta'}{dt} &=&  
\frac{m_{\rm p}}{m_{\rm t}} \left(\frac{d\beta}{dt}\right)_{{\rm t},B} 
\bigg|_{ \vec{r}_{\rm t} \rightarrow \frac{m_{\rm t}}{m_{\rm p}+m_{\rm t}} \vec{r}_{\rm t}  }^{(a,e,i,\Omega,\omega,f) \rightarrow (a',e',i',\Omega',\omega',f')   }
\nonumber
\\
&+&
\left(\frac{d\beta}{dt}\right)_{{\rm t},B}
\bigg|_{ \vec{r}_{\rm t} \rightarrow \frac{-m_{\rm p}}{m_{\rm p}+m_{\rm t}} \vec{r}_{\rm t}  }^{(a,e,i,\Omega,\omega,f) \rightarrow (a',e',i',\Omega',\omega',f')   }
\nonumber
\\
&-&
\left(\frac{m_{\rm p}+m_{\rm t}}{m_{\rm t}}\right)
\left(\frac{d\beta}{dt}\right)_{{\rm t},B}
\bigg|_{ \vec{r}_{\rm t} \rightarrow 0 }^{(a,e,i,\Omega,\omega,f) \rightarrow (a',e',i',\Omega',\omega',f')   }
.
\label{bary}
\end{eqnarray}

\begin{figure}
\centerline{
  \includegraphics[width=1.00\textwidth,height=7cm]{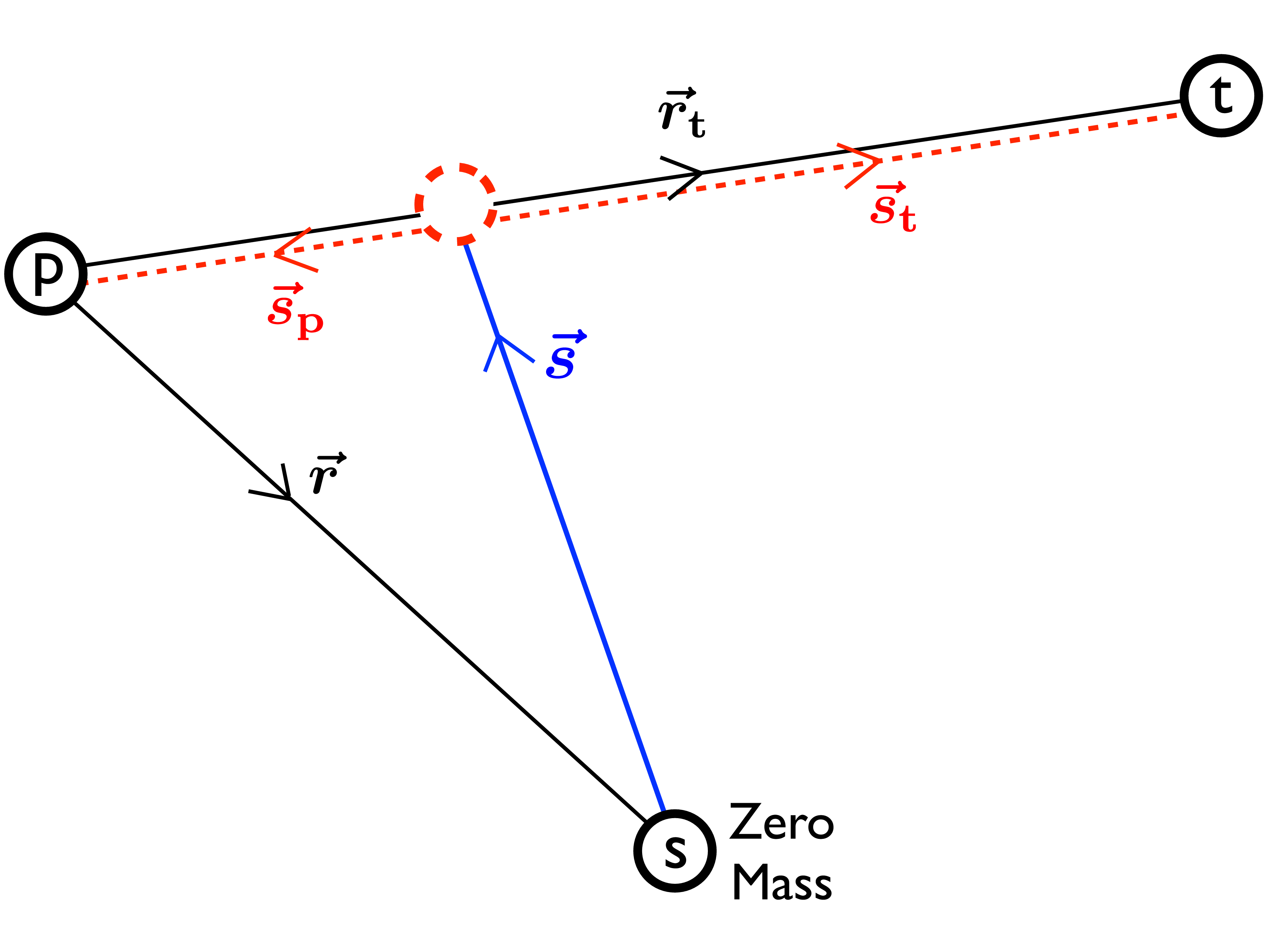}
}
\caption{Vectors used in the derivation of equation (\ref{bary}) to obtain the
equations of motion with orbital elements that are measured with respect to the 
centre of mass of the primary and tertiary.
}
\label{tridia}   
\end{figure}

I now write out this expression for each transformed orbital element through equations (\ref{rtBnear}), (\ref{dadtROT1})-(\ref{dodtROT1}) and (\ref{SjBa})-(\ref{SjBo}) first by rewriting the unprimed $B$ term expressions as

\begin{eqnarray}
\left(\frac{da}{dt}\right)_{{\rm t},B} &=&
  \frac{2Gm_{\rm t} S_{({\rm t},B,a)}}{n\sqrt{1-e^2}} 
\left( \frac{r_{\rm t} }{r_{{\rm t},B}^3} \right) 
\nonumber
\\
&=& \frac{2Gm_{\rm t} \left(- S_{({\rm t},A,a)} \alpha_{\rm sf} - e \sin{f}  \right)}{n\sqrt{1-e^2}r^2} 
\left(1 - \kappa_t \alpha_{\rm sf} + \alpha_{\rm sf}^2 \right)^{-\frac{3}{2}}
,
\label{dadtROT1sca}
\\
\left(\frac{de}{dt}\right)_{{\rm t},B} &=&
  \frac{Gm_{\rm t}\sqrt{1-e^2} S_{({\rm t},B,e)}}{2an\left(1 + e\cos{f}\right)} 
\left( \frac{r_{\rm t}}{r_{{\rm t},B}^3} \right) 
\nonumber
\\
&=& \frac{Gm_{\rm t}\sqrt{1-e^2} \left(- S_{({\rm t},A,e)}\alpha_{\rm sf} - 2 \sin{f} - e \sin{2f} \right)}{2an\left(1 + e\cos{f}\right)r^2}
\nonumber
\\
&\times&
\left(1 - \kappa_t \alpha_{\rm sf} + \alpha_{\rm sf}^2 \right)^{-\frac{3}{2}}
,
\label{dedtROT1sca}
\\
\left(\frac{di}{dt}\right)_{{\rm t},B} &=&
  \frac{Gm_{\rm t}\sqrt{1-e^2}S_{({\rm t},B,i)}}{an\left(1 + e\cos{f}\right)} 
\left( \frac{r_{\rm t}}{r_{{\rm t},B}^3} \right) 
\nonumber
\\
&=&
-\frac{Gm_{\rm t}\sqrt{1-e^2}S_{({\rm t},A,i)}\alpha_{\rm sf}}{an\left(1 + e\cos{f}\right)r^2} 
\left(1 - \kappa_t \alpha_{\rm sf} + \alpha_{\rm sf}^2 \right)^{-\frac{3}{2}}
,
\label{didtROT1sca}
\\
\left(\frac{d\Omega}{dt}\right)_{{\rm t},B} &=&
  \frac{Gm_{\rm t}\sqrt{1-e^2}S_{({\rm t},B,\Omega)}}{an\left(1 + e\cos{f}\right)}
 \left( \frac{r_{\rm t}}{r_{{\rm t},B}^3} \right) 
\nonumber
\\
&=&
-\frac{Gm_{\rm t}\sqrt{1-e^2}S_{({\rm t},A,\Omega)}\alpha_{\rm sf}}{an\left(1 + e\cos{f}\right)r^2}
\left(1 - \kappa_t \alpha_{\rm sf} + \alpha_{\rm sf}^2 \right)^{-\frac{3}{2}}
,
\label{dOdtROT1sca}
\\
\left(\frac{d\omega}{dt}\right)_{{\rm t},B} &=&
  \frac{Gm_{\rm t}\sqrt{1-e^2}S_{({\rm t},B,\omega)}}{2aen\left(1 + e\cos{f}\right)}
  \left( \frac{r_{\rm t}}{r_{{\rm t},B}^3} \right) 
\nonumber
\\
&=&
\frac{Gm_{\rm t}\sqrt{1-e^2}\left(-S_{({\rm t},A,\omega)}\alpha_{\rm sf} + 2 \cos{f} + 2 e \cos^2{f} \right)}{2aen\left(1 + e\cos{f}\right)r^2}
\nonumber
\\
&\times&
\left(1 - \kappa_t \alpha_{\rm sf} + \alpha_{\rm sf}^2 \right)^{-\frac{3}{2}}
.
\label{dodtROT1sca}
\end{eqnarray}

The forms in equations (\ref{dadtROT1sca})-(\ref{dodtROT1sca}) facilitate the derivation of the equations of motion, which are:

\begin{eqnarray}
\frac{da'}{dt} &=&
\frac{2Gm_{\rm p} \left(- S_{({\rm t},A,a')} \alpha_{\rm sf} \left(\frac{m_t}{m_p + m_t} \right) - e' \sin{f'}  \right)}{n'\sqrt{1-e'^2}r'^2} 
\nonumber
\\
&\times&
\left(1 - \kappa'_t \alpha_{\rm sf} \left(\frac{m_t}{m_p + m_t} \right) + \alpha_{\rm sf}^2 \left(\frac{m_t}{m_p + m_t} \right)^2\right)^{-3/2}
\nonumber
\\
&+& 
\frac{2Gm_{\rm t} \left( S_{({\rm t},A,a')} \alpha_{\rm sf} \left(\frac{m_p}{m_p + m_t} \right) - e' \sin{f'}  \right)}{n'\sqrt{1-e'^2}r'^2} 
\nonumber
\\
&\times&
\left(1 + \kappa'_t \alpha_{\rm sf} \left(\frac{m_p}{m_p + m_t} \right) + \alpha_{\rm sf}^2 \left(\frac{m_p}{m_p + m_t} \right)^2\right)^{-3/2}
\nonumber
\\
&-&
\frac{2G\left(m_{\rm p}+m_{\rm t}\right) \left( - e' \sin{f'}  \right)}{n'\sqrt{1-e'^2}r'^2} 
\label{dadtROT2sca}
\\
\frac{de'}{dt} &=&
\frac{Gm_{\rm p}\sqrt{1-e'^2} \left(- S_{({\rm t},A,e')}\alpha_{\rm sf}\left(\frac{m_t}{m_p + m_t} \right)  - 2 \sin{f'} - e' \sin{2f'} \right)}{2a'n'\left(1 + e'\cos{f'}\right)r'^2}
\nonumber
\\
&\times&
\left(1 - \kappa'_t \alpha_{\rm sf}\left(\frac{m_t}{m_p + m_t} \right)  + \alpha_{\rm sf}^2 \left(\frac{m_t}{m_p + m_t} \right)^2 \right)^{-3/2}
\nonumber
\\
&+&
\frac{Gm_{\rm t}\sqrt{1-e'^2} \left(S_{({\rm t},A,e')}\alpha_{\rm sf}\left(\frac{m_p}{m_p + m_t} \right)  - 2 \sin{f'} - e' \sin{2f'} \right)}{2a'n'\left(1 + e'\cos{f'}\right)r'^2}
\nonumber
\\
&\times&
\left(1 + \kappa'_t \alpha_{\rm sf}\left(\frac{m_p}{m_p + m_t} \right)  + \alpha_{\rm sf}^2 \left(\frac{m_p}{m_p + m_t} \right)^2 \right)^{-3/2}
\nonumber
\\
&-&
\frac{G\left(m_{\rm p}+m_{\rm t}\right)\sqrt{1-e'^2} \left(- 2 \sin{f'} - e' \sin{2f'} \right)}{2a'n'\left(1 + e'\cos{f'}\right)r'^2}
,
\label{dedtROT2sce}
\\
\frac{di'}{dt} &=&
-\frac{Gm_{\rm t}m_{\rm p}\sqrt{1-e'^2}S_{({\rm t},A,i')}\alpha_{\rm sf}}{a'n'\left(1 + e'\cos{f'}\right)r'^2\left(m_p + m_t\right)} 
\nonumber
\\
&\times&
\left(1 - \kappa'_t \alpha_{\rm sf}\left(\frac{m_t}{m_p + m_t} \right) + \alpha_{\rm sf}^2\left(\frac{m_t}{m_p + m_t} \right)^2 \right)^{-3/2}
\nonumber
\\
&+&
\frac{Gm_{\rm t}m_{\rm p}\sqrt{1-e'^2}S_{({\rm t},A,i')}\alpha_{\rm sf}}{a'n'\left(1 + e'\cos{f'}\right)r'^2\left(m_p + m_t\right)} 
\nonumber
\\
&\times&
\left(1 + \kappa'_t \alpha_{\rm sf}\left(\frac{m_p}{m_p + m_t} \right) + \alpha_{\rm sf}^2\left(\frac{m_p}{m_p + m_t} \right)^2 \right)^{-3/2}
,
\label{didtROT2sci}
\\
\frac{d\Omega'}{dt} &=&
-\frac{Gm_{\rm t}m_{\rm p}\sqrt{1-e'^2}S_{({\rm t},A,\Omega')}\alpha_{\rm sf}}{a'n'\left(1 + e'\cos{f'}\right)r'^2\left(m_p + m_t\right)}
 \nonumber
\\
&\times&
\left(1 - \kappa'_t \alpha_{\rm sf}\left(\frac{m_t}{m_p + m_t} \right) + \alpha_{\rm sf}^2\left(\frac{m_t}{m_p + m_t} \right)^2 \right)^{-3/2}
\nonumber
\\
&+&
\frac{Gm_{\rm t}m_{\rm p}\sqrt{1-e'^2}S_{({\rm t},A,\Omega')}\alpha_{\rm sf}}{a'n'\left(1 + e'\cos{f'}\right)r'^2\left(m_p + m_t\right)} 
\nonumber
\\
&\times&
\left(1 + \kappa'_t \alpha_{\rm sf}\left(\frac{m_p}{m_p + m_t} \right) + \alpha_{\rm sf}^2\left(\frac{m_p}{m_p + m_t} \right)^2 \right)^{-3/2}
,
\label{dOdtROT2scO}
\\
\frac{d\omega'}{dt} &=&
\frac{Gm_{\rm p}\sqrt{1-e'^2}\left(-S_{({\rm t},A,\omega')}\alpha_{\rm sf}\left(\frac{m_t}{m_p + m_t} \right) + 2 \cos{f'} + 2 e' \cos^2{f'} \right)}{2a'e'n'\left(1 + e'\cos{f'}\right)r'^2}
\nonumber
\\
&\times&
\left(1 - \kappa'_t \alpha_{\rm sf} \left(\frac{m_t}{m_p + m_t} \right) + \alpha_{\rm sf}^2 \left(\frac{m_t}{m_p + m_t} \right)^2 \right)^{-3/2}
\nonumber
\\
&+& \frac{Gm_{\rm t}\sqrt{1-e'^2}\left(S_{({\rm t},A,\omega')}\alpha_{\rm sf}\left(\frac{m_p}{m_p + m_t} \right) + 2 \cos{f'} + 2 e' \cos^2{f'} \right)}{2a'e'n'\left(1 + e'\cos{f'}\right)r'^2}
\nonumber
\\
&\times&
\left(1 + \kappa'_t \alpha_{\rm sf} \left(\frac{m_p}{m_p + m_t} \right) + \alpha_{\rm sf}^2 \left(\frac{m_p}{m_p + m_t} \right)^2 \right)^{-3/2}
\nonumber
\\
&-&
\frac{G\left(m_{\rm p}+m_{\rm t}\right)\sqrt{1-e'^2}\left(2 \cos{f'} + 2 e' \cos^2{f'} \right)}{2a'e'n'\left(1 + e'\cos{f'}\right)r'^2}
,
\label{dodtROT2sco}
\\
\frac{df'}{dt} &=& \frac{n' \left(1 + e' \cos{f'}\right)^2}{\left(1 - e'^2\right)^{3/2}}
                - \frac{d\omega'}{dt} - \cos{i'} \frac{d\Omega'}{dt}
.
\label{dodtROT2scf}
\end{eqnarray}

\noindent{}Equations (\ref{dadtROT2sca})-(\ref{dodtROT2scf}) are the complete equations of motion in orbital elements for a secondary's orbit with respect to the barycentre of the primary and tertiary.  Note that the orbital elements in the $S$ variables are primed.

\subsection{Preparations for averaging}

By inspection, one may note that all terms to zeroth order in $\alpha_{\rm sf}$ for all of the equations cancel.  Therefore, the time evolution of the elements are always dependent on the ratio of semimajor axes.  When expressed as a power series in semimajor axis ratio, the equations of motion become

\begin{eqnarray}
\frac{da'}{dt} &=& 
-\frac{2G S_{({\rm t},A,a')} \alpha_{\rm sf}}{n'\sqrt{1-e'^2}r'^2}
\left(\frac{m_{\rm p}m_{\rm t}}{m_{\rm p} + m_{\rm t}}\right)
\nonumber
\\
&\times&
 \sum_{u = 0}^{\infty} 
     \left(
     \begin{array}{c}
     \frac{1}{2} + u \\
     \frac{1}{2}
     \end{array}
     \right)
     \left(\frac{\alpha_{\rm sf}}{m_{\rm p}+m_{\rm t}}\right)^u 
\bigg\lbrace
     m_{\rm t}^u \left[\kappa'_{\rm t} - \left(\frac{m_{\rm t}}{m_{\rm p}+m_{\rm t}}\right) \alpha_{\rm sf}\right]^u
\nonumber
\\
&-& 
     \left(-m_{\rm p}\right)^u \left[\kappa'_{\rm t} + \left(\frac{m_{\rm p}}{m_{\rm p}+m_{\rm t}}\right) \alpha_{\rm sf}\right]^u
\bigg\rbrace
\nonumber
\\
&-&
\frac{2G e' \sin{f'}}{n'\sqrt{1-e'^2}r'^2}
\nonumber
\\
&\times&
 \sum_{u = 1}^{\infty}
     \left(
     \begin{array}{c}
     \frac{1}{2} + u \\
     \frac{1}{2}
     \end{array}
     \right)
     \left(\frac{\alpha_{\rm sf}}{m_{\rm p}+m_{\rm t}}\right)^u 
\bigg\lbrace
     m_p m_{\rm t}^u \left[\kappa'_{\rm t} - \left(\frac{m_{\rm t}}{m_{\rm p}+m_{\rm t}}\right) \alpha_{\rm sf}\right]^u
\nonumber
\\
&+& 
     m_t \left(-m_{\rm p}\right)^u \left[\kappa'_{\rm t} + \left(\frac{m_{\rm p}}{m_{\rm p}+m_{\rm t}}\right) \alpha_{\rm sf}\right]^u
\bigg\rbrace
,
\\
&&
\nonumber
\\
&&
\nonumber
\\
\frac{de'}{dt} &=& 
-\frac{G S_{({\rm t},A,e')} \sqrt{1-e'^2} \alpha_{\rm sf}}{2a'n'\left(1 + e' \cos{f'} \right) r'^2}
\left(\frac{m_{\rm p}m_{\rm t}}{m_{\rm p} + m_{\rm t}}\right)
\nonumber
\\
&\times&
 \sum_{u = 0}^{\infty} 
     \left(
     \begin{array}{c}
     \frac{1}{2} + u \\
     \frac{1}{2}
     \end{array}
     \right)
     \left(\frac{\alpha_{\rm sf}}{m_{\rm p}+m_{\rm t}}\right)^u 
\bigg\lbrace
     m_{\rm t}^u \left[\kappa'_{\rm t} - \left(\frac{m_{\rm t}}{m_{\rm p}+m_{\rm t}}\right) \alpha_{\rm sf}\right]^u
\nonumber
\\
&-&
     \left(-m_{\rm p}\right)^u \left[\kappa'_{\rm t} + \left(\frac{m_{\rm p}}{m_{\rm p}+m_{\rm t}}\right) \alpha_{\rm sf}\right]^u
\bigg\rbrace
\nonumber
\\
&-&
\frac{G \sqrt{1-e'^2} \left(2\sin{f'} + e' \sin{2f'}\right)}{2a'n'\left(1 + e' \cos{f'} \right)r'^2}
\nonumber
\\
&\times&
 \sum_{u = 1}^{\infty}
     \left(
     \begin{array}{c}
     \frac{1}{2} + u \\
     \frac{1}{2}
     \end{array}
     \right)
     \left(\frac{\alpha_{\rm sf}}{m_{\rm p}+m_{\rm t}}\right)^u 
\bigg\lbrace
     m_p m_{\rm t}^u \left[\kappa'_{\rm t} - \left(\frac{m_{\rm t}}{m_{\rm p}+m_{\rm t}}\right) \alpha_{\rm sf}\right]^u
\nonumber
\\
&+& 
     m_t \left(-m_{\rm p}\right)^u \left[\kappa'_{\rm t} + \left(\frac{m_{\rm p}}{m_{\rm p}+m_{\rm t}}\right) \alpha_{\rm sf}\right]^u
\bigg\rbrace
,
\\
&&
\nonumber
\\
&&
\nonumber
\\
\frac{di'}{dt} &=& 
-\frac{G S_{({\rm t},A,i')} \sqrt{1-e'^2} \alpha_{\rm sf}}{a'n'\left(1 + e' \cos{f'} \right) r'^2 } 
\left(\frac{m_{\rm p}m_{\rm t}}{m_{\rm p} + m_{\rm t}}\right)
\nonumber
\\
&\times&
 \sum_{u = 0}^{\infty} 
     \left(
     \begin{array}{c}
     \frac{1}{2} + u \\
     \frac{1}{2}
     \end{array}
     \right)
     \left(\frac{\alpha_{\rm sf}}{m_{\rm p}+m_{\rm t}}\right)^u 
\bigg\lbrace
     m_{\rm t}^u \left[\kappa'_{\rm t} - \left(\frac{m_{\rm t}}{m_{\rm p}+m_{\rm t}}\right) \alpha_{\rm sf}\right]^u
\nonumber
\\
&-& 
     \left(-m_{\rm p}\right)^u \left[\kappa'_{\rm t} + \left(\frac{m_{\rm p}}{m_{\rm p}+m_{\rm t}}\right) \alpha_{\rm sf}\right]^u
\bigg\rbrace
,
\\
&&
\nonumber
\\
&&
\nonumber
\\
\frac{d\Omega'}{dt} &=& 
-\frac{G S_{({\rm t},A,\Omega')} \sqrt{1-e'^2} \alpha_{\rm sf}}{a'n'\left(1 + e' \cos{f'} \right) r'^2} 
\left(\frac{m_{\rm p}m_{\rm t}}{m_{\rm p} + m_{\rm t}}\right)
\nonumber
\\
&\times&
 \sum_{u = 0}^{\infty} 
     \left(
     \begin{array}{c}
     \frac{1}{2} + u \\
     \frac{1}{2}
     \end{array}
     \right)
     \left(\frac{\alpha_{\rm sf}}{m_{\rm p}+m_{\rm t}}\right)^u 
\bigg\lbrace
     m_{\rm t}^u \left[\kappa'_{\rm t} - \left(\frac{m_{\rm t}}{m_{\rm p}+m_{\rm t}}\right) \alpha_{\rm sf}\right]^u
\nonumber
\\
&-& 
     \left(-m_{\rm p}\right)^u \left[\kappa'_{\rm t} + \left(\frac{m_{\rm p}}{m_{\rm p}+m_{\rm t}}\right) \alpha_{\rm sf}\right]^u
\bigg\rbrace
,
\\
&&
\nonumber
\\
&&
\nonumber
\\
\frac{d\omega'}{dt} &=& 
-\frac{G S_{({\rm t},A,\omega')} \sqrt{1-e'^2} \alpha_{\rm sf}}{2a'e'n'\left(1 + e' \cos{f'} \right) r'^2 } 
\left(\frac{m_{\rm p}m_{\rm t}}{m_{\rm p} + m_{\rm t}}\right)
\nonumber
\\
&\times&
 \sum_{u = 0}^{\infty} 
     \left(
     \begin{array}{c}
     \frac{1}{2} + u \\
     \frac{1}{2}
     \end{array}
     \right)
     \left(\frac{\alpha_{\rm sf}}{m_{\rm p}+m_{\rm t}}\right)^u 
\bigg\lbrace
     m_{\rm t}^u \left[\kappa'_{\rm t} - \left(\frac{m_{\rm t}}{m_{\rm p}+m_{\rm t}}\right) \alpha_{\rm sf}\right]^u
\nonumber
\\
&-& 
     \left(-m_{\rm p}\right)^u \left[\kappa'_{\rm t} + \left(\frac{m_{\rm p}}{m_{\rm p}+m_{\rm t}}\right) \alpha_{\rm sf}\right]^u
\bigg\rbrace
\nonumber
\\
&+&
\frac{G \sqrt{1-e'^2} \left(2\cos{f'} + 2e' \cos^2{f'}\right)}{2a'e'n'\left(1 + e' \cos{f'} \right)r'^2}
\nonumber
\\
&\times&
 \sum_{u = 1}^{\infty}
     \left(
     \begin{array}{c}
     \frac{1}{2} + u \\
     \frac{1}{2}
     \end{array}
     \right)
     \left(\frac{\alpha_{\rm sf}}{m_{\rm p}+m_{\rm t}}\right)^u 
\bigg\lbrace
     m_p m_{\rm t}^u \left[\kappa'_{\rm t} - \left(\frac{m_{\rm t}}{m_{\rm p}+m_{\rm t}}\right) \alpha_{\rm sf}\right]^u
\nonumber
\\
&+& 
     m_t \left(-m_{\rm p}\right)^u \left[\kappa'_{\rm t} + \left(\frac{m_{\rm p}}{m_{\rm p}+m_{\rm t}}\right) \alpha_{\rm sf}\right]^u
\bigg\rbrace
.
\end{eqnarray}

\noindent{}Note that the starting index on the first summation may be increased to 1 because the $u = 0$ term vanishes.  Also,

\begin{equation}
\frac{df'}{dt} = \frac{n' \left(1 + e' \cos{f'}\right)^2}{\left(1 - e'^2\right)^{3/2}}
                - \frac{d\omega'}{dt} - \cos{i'} \frac{d\Omega'}{dt}
.
\end{equation}

Regarding the fully coplanar equations of motion, by analogy with the transition from the general to the planar case in the reference frame of the primary, I can write

\begin{eqnarray}
\left(\frac{da'}{dt}\right)^{\rm P} &=& \frac{da'}{dt} \bigg|_{S_{({\rm t},A,a')} \rightarrow S_{({\rm t},A,a')}^{\rm P}}^{\kappa'_{\rm t} \rightarrow \kappa_{\rm t}^{'\rm P}} 
,
\\
\left(\frac{de'}{dt}\right)^{\rm P} &=& \frac{de'}{dt} \bigg|_{S_{({\rm t},A,e')} \rightarrow S_{({\rm t},A,e')}^{\rm P}}^{\kappa'_{\rm t} \rightarrow \kappa_{\rm t}^{'\rm P}} 
,
\\
\left(\frac{d\varpi'}{dt}\right)^{\rm P} &=& \frac{d\omega'}{dt} \bigg|_{S_{({\rm t},A,\omega')} \rightarrow S_{({\rm t},A,\varpi')}^{\rm P}}^{\kappa'_{\rm t} \rightarrow \kappa_{\rm t}^{'\rm P}} 
.
\end{eqnarray}

\section{Averaging Procedure}  \label{avgsec}

Until now, all the equations of motion describe
how the secondary orbit changes throughout every
revolution or flyby of every body in the system.  
Sometimes, however, the secondary is far away enough
from some of the massive bodies such that their
gravitational influence produces small oscillations of the
secondary orbit.  These oscillations may be averaged
over any or all of the orbits to yield a net change
in orbital elements.  Averaged quantities are particularly
useful to determine the long-term (often referred to as 
{\it secular}) evolution of a dynamical system, and provide
fundamental insights that may be lost in the detail
of the full, unaveraged equations.

I perform averaging only for $N=3$ body systems, but
consider every combination of averaging for these systems.
For example, if the secondary is close to the primary but far from
the tertiary, then one could average over only the secondary orbit
or both orbits.  The usefulness of either approach is dependent
on the timescale for change sought, and the details of the
system studied.

I denote the orbit average of  
an arbitrary variable $\beta$ with
a hat or a tilde such that averaging over
the secondary and tertiary respectively is 
expressed as

\begin{equation}
\widehat{\frac{d\beta}{dt}} \equiv \frac{n}{2\pi}
\int_{0}^{2\pi} \frac{d\beta}{dt} \frac{dt}{df}df 
,
\
\
\
\widetilde{\frac{d\beta}{dt}} \equiv \frac{n_{\rm t}}{2\pi}
\int_{0}^{2\pi} \frac{d\beta}{dt} \frac{dt}{df_{\rm t}}df_{\rm t} 
\end{equation}

\noindent{}and similarly for $\beta'$, with 

\begin{equation}
\frac{dt}{df} = \frac{\left(1 - e^2\right)^{3/2}}{n \left(1 + e \cos{f}\right)^2}
,
\
\
\
\frac{dt}{df_{\rm t}} = 
\frac{\left(1 - e_{\rm t}^2\right)^{3/2}}{n_{\rm t} \left(1 + e_{\rm t} \cos{f_{\rm t}}\right)^2}
.
\end{equation}

The averaging procedure requires me to perform
integrals which are difficult, if not impossible,
to solve analytically with the equations of motion
in their full generality.  Therefore, I must
make an approximation.  So I utilize the power series
representations generated from equation (\ref{series}) and 
assume that either
$\alpha_{\rm sf} \ll 1$ or $\alpha_{\rm sc} \ll 1$.
My auxiliary variables are already written in forms 
to isolate $f$ and $f_t$, facilitating the computation.

I use the algebraic manipulation
software package {\it Mathematica} to perform the
averaging, but must do so on a term-by-term
basis.  I find that the most expeditious procedure
is to precompute individual integrals symbolically 
where the integrand is a function of $\left(1 + e \cos{f}\right)^q$
or $\left(1 + e' \cos{f'}\right)^q$, 
where $q$ is an integer.  The result is in terms of hypergeometric 
functions of $q$.  These symbolic solutions can
then be used when computing coefficients to different
orders.  The integrals are also most easily computed
when all powers of $\sin{f}$, $\cos{f}$, $\sin{f'}$, and
$\cos{f'}$ are broken down into single powers through 
multiple-angle formulae.

I report final results to selected orders of
powers of distances or semimajor axes
depending on the length of
the expressions.  Often, double averaging yields much
simpler formula than single averaging.  However,
the singly-averaged formulas may be important
depending on the timescales considered.  I compute
averages for both the general equations of motion
in the rotated frame and the equations of motion when
all three bodies are coplanar. 

\section{Averaging when secondary is closer than tertiary} \label{scsec}

Here the relevant distance ratio is $\alpha_{\rm sc}$ and the orbital elements
are unprimed, meaning that they are measured with respect to the primary.

\subsection{Averaging over secondary orbit only}

The tertiary's orbit here may be eccentric, parabolic or hyperbolic.  In the
latter two cases, computing the resulting change in the secondary's orbital
parameters may be particularly useful if the impulse approximation is not
applicable.

\subsubsection{Nonplanar equations}

\begin{eqnarray}
\widehat{\left( \frac{da}{dt} \right)}_{\rm sc} 
&=&
0 
\label{Gurf1}
\\
\widehat{\left( \frac{de}{dt} \right)}_{\rm sc} &=&
\left(\frac{1}{r_{\rm t}^3}\right)
\frac{15 Gm_{\rm t}e\sqrt{1 - e^2}}{2n} D_{{\rm t},1} D_{{\rm t},2}
+
\mathcal{O}\left(
\frac{Gm_{\rm t}}{n}
\frac{a}{r_{\rm t}^4} \right)
,
\\
\widehat{\left( \frac{di}{dt} \right)}_{\rm sc} &=&
\left(\frac{1}{r_{\rm t}^3}\right)
\frac{3 Gm_{\rm t}\sin{i} \sin{\chi_{\rm t}}}{2n\sqrt{1 - e^2}} 
\nonumber
\\
&\times&
\left[ 
D_{{\rm t},1} \left(1 - e^2\right) \sin{\omega} 
+ 
D_{{\rm t},2} \left(1 + 4e^2\right) \cos{\omega} 
\right]
\nonumber
\\
&+&
\mathcal{O}\left(
\frac{Gm_{\rm t}}{n}
\frac{a}{r_{\rm t}^4} \right)
,
\\
\widehat{\left( \frac{d\Omega}{dt} \right)}_{\rm sc} &=&
\left(\frac{1}{r_{\rm t}^3}\right)
\frac{3 Gm_{\rm t}\sin{\chi_{\rm t}}}{2n\sqrt{1 - e^2}} 
\nonumber
\\
&\times&
\left[ 
-D_{{\rm t},1} \left(1 - e^2\right) \cos{\omega} 
+ 
D_{{\rm t},2} \left(1 + 4e^2\right) \sin{\omega} 
\right]
\nonumber
\\
&+&
\mathcal{O}\left(
\frac{Gm_{\rm t}}{n}
\frac{a}{r_{\rm t}^4} \right)
,
\\
\widehat{\left( \frac{d\omega}{dt} \right)}_{\rm sc} &=&
\left(\frac{1}{r_{\rm t}^3}\right)
\frac{3 Gm_{\rm t}}{2n\sqrt{1 - e^2}} 
\big[ 
5 D_{{\rm t},2}^2 - 1 + e^2 
-D_{{\rm t},1}^2 \left(1 - e^2\right)
\nonumber
\\
&-&
D_{{\rm t},2} \left(1 + 4e^2\right) \cos{\omega}  \cos{\chi_{\rm t}}
\nonumber
\\
&+&
D_{{\rm t},1} \left(1 - e^2\right) \cos{i} \cos{\omega} \sin{\chi_{\rm t}}
\big] 
+
\mathcal{O}\left(
\frac{Gm_{\rm t}}{n}
\frac{a}{r_{\rm t}^4} \right)
.
\label{Gurf2}
\end{eqnarray}

As a check on equations (\ref{Gurf1}-\ref{Gurf2}), I consider the
expressions for the motion of a Martian satellite by \cite{guretal2007}.
Those authors produce similar singly-averaged expressions, but for
a satellite (secondary) of Mars (primary), which orbits the Sun (tertiary).
The leading order terms in my equations (\ref{Gurf1}-\ref{Gurf2}) correctly reduce to their
equations (38a-38e) under their assumption that 
the Martian orbit around the Sun is circular.  Also, under this assumption,
their $\tilde{\Omega}$ (their equation 41) is equivalent to my 
$\chi_{\rm t}$ (my equation \ref{chit}).

\subsubsection{Coplanar equations}

\begin{eqnarray}
{\widehat{\left( \frac{da}{dt} \right)}^{\rm P}}_{\rm sc}
&=&
0
\\
{\widehat{\left( \frac{de}{dt} \right)}^{\rm P}}_{\rm sc}
&=&
\left(\frac{1}{r_{\rm t}^3}\right)
\frac{15 Gm_{\rm t}e\sqrt{1 - e^2}}{2n} D_{{\rm t},1}^{\rm P} D_{{\rm t},2}^{\rm P}
\nonumber
\\
&+&
\left(\frac{a}{r_{\rm t}^4}\right)
\frac{15 Gm_{\rm t}\sqrt{1 - e^2}}{16n} D_{{\rm t},1}^{\rm P}
\nonumber
\\
&\times&
\left[4 + 3e^2 - 5\left(D_{{\rm t},1}^{\rm P}\right)^2 \left(1 - e^2\right) - 5\left(D_{{\rm t},2}^{\rm P}\right)^2 \left(1 + 6e^2\right) \right]
\nonumber
\\
&+& \mathcal{O}\left(
\frac{Gm_{\rm t}}{n}
\frac{a^2}{r_{\rm t}^5} \right)
,
\\
{\widehat{\left( \frac{d\varpi}{dt} \right)}^{\rm P}}_{\rm sc}
&=&
-
\left(\frac{1}{r_{\rm t}^3}\right)
\frac{3 Gm_{\rm t}\sqrt{1 - e^2}}{2n} 
\left[
1
+
\left(D_{{\rm t},1}^{\rm P}\right)^2
-
4 \left(D_{{\rm t},2}^{\rm P}\right)^2
\right]
\nonumber
\\
&+&
\left(\frac{a}{r_{\rm t}^4}\right)
\frac{15 Gm_{\rm t}\sqrt{1 - e^2}}{16en} 
D_{{\rm t},2}^{\rm P}
\nonumber
\\
&\times&
\left[
4 + 9 e^2 
- 5 \left(D_{{\rm t},1}^{\rm P}\right)^2
\left(1 - 3e^2 \right)
- 5 \left(D_{{\rm t},2}^{\rm P}\right)^2
\left(1 + 4e^2 \right)
\right]
\nonumber
\\
&+&
\mathcal{O}\left(
\frac{Gm_{\rm t}}{n}
\frac{a^2}{r_{\rm t}^5} \right)
.
\end{eqnarray}

Although the averaged semimajor axis remains fixed, the eccentricity does not.  If the eccentricity varies enough, then the secondary may collide with the primary or escape the system during a single orbit or flyby of the tertiary.

\subsection{Averaging over both orbits}

These equations should be used when one seeks the very long term evolution (over many tertiary orbits) of the system.

\subsubsection{Nonplanar equations}

Here I assume that the tertiary is on a bound (elliptical) orbit.  Then

\begin{eqnarray}
\widetilde{\widehat{\left( \frac{da}{dt} \right)}}_{\rm sc} 
&=&
0
,
\\
\widetilde{\widehat{\left( \frac{de}{dt} \right)}}_{\rm sc} 
&=&
\left(\frac{1}{a_{\rm t}^3}\right)
\frac{15Gm_{\rm t} e \sqrt{1 - e^2} \sin{2\omega} \sin^2{i}}
{8 n \left(1 - e_{\rm t}^2 \right)^{3/2}}
\nonumber
\\
&-&
\left(\frac{a}{a_{\rm t}^4}\right)
\frac{15Gm_{\rm t} e_{\rm t}\sqrt{1-e^2}}{512 n \left(1-e_{\rm t}^2\right)^{5/2}} 
\nonumber
\\
&\times&
\bigg\lbrace
\cos \left(\Omega -\varpi_{\rm t}\right) \big(210 e^2 \sin^2{i} \sin (3 \omega )
\nonumber
\\
&+&
\left(3 e^2+4\right) (5 \cos (2 i)+3) \sin{\omega}\big)
\nonumber
\\
&+&
2 \sin \left(\Omega -\varpi_{\rm t}\right) \cos{i} \cos{\omega}  \big[7 \left(30 e^2 \sin^2{i} \cos (2 \omega )-9 e^2-2\right)
\nonumber
\\
&+&
15 \left(5 e^2+2\right) \cos (2 i)\big]
\bigg\rbrace
+
\mathcal{O}\left(
\frac{Gm_{\rm t}}{n}
\frac{a^2}{a_{\rm t}^5} \right)
\label{Koze}
,
\\
\widetilde{\widehat{\left( \frac{di}{dt} \right)}}_{\rm sc} 
&=&
-\left(\frac{1}{a_{\rm t}^3}\right)
\frac{15Gm_{\rm t} e^2 \sin{2\omega} \sin{2i}}
{16 n \sqrt{1 - e^2} \left(1 - e_{\rm t}^2 \right)^{3/2}}
\nonumber
\\
&+&
\left(\frac{a}{a_{\rm t}^4}\right)
\frac{15 G m_{\rm t} e e_{\rm t} \sin{i}  }{256 n\sqrt{1-e^2} \left(1-e_{\rm t}^2\right)^{5/2}}
\nonumber
\\
&\times&
\bigg\lbrace
20 \cos \left(\Omega -\varpi_{\rm t}\right) \cos{i} \sin{\omega} \left(7 e^2 \cos (2 \omega )+5 e^2+2\right)
\nonumber
\\
&-&
\sin \left(\Omega -\varpi_{\rm t}\right) \cos{\omega} \big[-35 e^2 (3 \cos (2 i)+1) \cos (2 \omega )
\nonumber
\\
&+&
15 \left(5 e^2+2\right) \cos (2 i)+37 e^2+26\big]
\bigg\rbrace
+
\mathcal{O}\left(
\frac{Gm_{\rm t}}{n}
\frac{a^2}{a_{\rm t}^5} \right)
,
\\
\widetilde{\widehat{\left( \frac{d\Omega}{dt} \right)}}_{\rm sc} 
&=&
\left(\frac{1}{a_{\rm t}^3}\right)
\frac{3Gm_{\rm t} \cos{i} \left(-2 - 3e^2 + 5e^2 \cos{2\omega} \right)}
{8 n \sqrt{1 - e^2} \left(1 - e_{\rm t}^2 \right)^{3/2}}
\nonumber
\\
&+&
\left(\frac{a}{a_{\rm t}^4}\right)
\frac{15 G m_{\rm t} e e_{\rm t}}{256 n \sqrt{1-e^2} \left(1-e_j^2\right)^{5/2}}
\nonumber
\\
&\times&
\bigg\lbrace
20 \cos \left(\Omega -\varpi_{\rm t}\right) \cos{i} \cos {\omega} \left(-7 e^2 \cos (2 \omega )+5 e^2+2\right)
\nonumber
\\
&-&
\sin \left(\Omega -\varpi_{\rm t}\right) \sin{\omega} \big[-35 e^2 (3 \cos (2 i)+1) \cos (2 \omega )
\nonumber
\\
&+&
15 \left(e^2+6\right) \cos (2 i)+17 e^2+46\big]
\bigg\rbrace
+
\mathcal{O}\left(
\frac{Gm_{\rm t}}{n}
\frac{a^2}{a_{\rm t}^5} \right)
,
\\
\widetilde{\widehat{\left( \frac{d\omega}{dt} \right)}}_{\rm sc} 
&=&
\left(\frac{1}{a_{\rm t}^3}\right)
\frac{3Gm_{\rm t} 
\left[
2 \left(1 - e^2\right) - 5 \sin^2{\omega} \left(\sin^2{i} - e^2\right)
\right]
}
{4 n \sqrt{1 - e^2} \left(1 - e_{\rm t}^2 \right)^{3/2}}
\nonumber
\\
&+&
\left(\frac{a}{a_{\rm t}^4}\right)
\frac{15 G m_{\rm t} e_{\rm t}}{1024 e n \sqrt{1-e^2} \left(1-e_{\rm t}^2\right)^{5/2}}
\bigg\lbrace
-4 \cos \left(\Omega -\varpi_{\rm t}\right)\cos{\omega} 
\nonumber
\\
&\times&
\big[89 e^4+35 e^2 \cos (2 \omega ) \left(\left(e^2-3\right) \cos (2 i)-5 e^2+3\right)
\nonumber
\\
&-& 
25 e^2+5 \left(-5 e^4+17 e^2+2\right) \cos (2 i)+6\big] 
\nonumber
\\
&-&
2  \sin \left(\Omega -\varpi_{\rm t}\right) \sin {\omega} 
\nonumber
\\
&\times&
\big[\cos{i} \left(8 e^4+70 e^2 \cos (2 \omega ) \left(4 e^2+3 \cos (2 i)-3\right)-237 e^2-2\right)
\nonumber
\\
&-&
15 \left(5 e^2+2\right) \cos (3 i)\big]
\bigg\rbrace +
\mathcal{O}\left(
\frac{Gm_{\rm t}}{n}
\frac{a^2}{a_{\rm t}^5} \right)
.
\label{Kozw}
\end{eqnarray}

\noindent{}Equations (\ref{Koze})-(\ref{Kozw}) are particularly important
because the leading order term in each equation is the foundation of
Lidov-Kozai theory \citep{lidov1961,kozai1962}, which was originally derived
in the test particle limit.  These terms are equivalent to equations (9.34) of \cite{valkar2006} when
their reduced mass of the primary and secondary is equal to unity,
as they do not consider the test particle limit.  \cite{guretal2007} do consider the test
particle limit, and their equations (42a-42e) match the leading order terms in my equations (\ref{Koze})-(\ref{Kozw})
in their limit of a circular tertiary orbit ($e_{\rm t} = 0$).

I also included the next, ($a/a_{\rm t}^4$) terms (often refereed to as the {\it octupole} terms) in full 
because of the interest they have recently attracted
in the astronomical community.  In particular, they can quantitatively affect
classic Lidov-Kozai dynamics, a fact previously missed because of the premature 
elimination of the nodes in a Hamiltonian derivation \citep[see the summary in][]{naoetal2013}.
Here, these terms are derived without 
appealing to Delaunay variables, and explicitly demonstrate how the evolution 
becomes dependent on the longitude of ascending node.  Note that 
all of these ($a/a_{\rm t}^4$) terms vanish when the tertiary is on
a circular orbit.

\subsubsection{Coplanar equations}

The planar versions of these terms are

\begin{eqnarray}
{\widetilde{\widehat{\left( \frac{da}{dt} \right)}}^{\rm P}}_{\rm sc}
&=& 0
,
\label{doubavga}
\\
{\widetilde{\widehat{\left( \frac{de}{dt} \right)}}^{\rm P}}_{\rm sc} 
&=& 0 \times \left(\frac{1}{a_{\rm t}^3} \right)
-
\left(\frac{a}{a_{\rm t}^4} \right)
\frac{15Gm_{\rm t} e_{\rm t}\sqrt{1 - e^2} \left(4 + 3 e^2\right)}
{64 n \left(1 - e_{\rm t}^2\right)^{5/2}}
\sin{\left(\varpi - \varpi_{\rm t}\right)}
\nonumber
\\
&+&
\left(\frac{a^2}{a_{\rm t}^5} \right)
\frac{315Gm_{\rm t} e e_{\rm t}^2\sqrt{1 - e^2} \left(2 + e^2\right)}
{256 n \left(1 - e_{\rm t}^2\right)^{7/2}}
\sin{\left[2\left(\varpi - \varpi_{\rm t}\right)\right]}
\nonumber
\\
&+&
\mathcal{O}\left(
\frac{Gm_{\rm t}}{n}
\frac{a^3}{a_{\rm t}^6} \right)
,
\\
{\widetilde{\widehat{\left( \frac{d\varpi}{dt} \right)}}^{\rm P}}_{\rm sc} 
&=& 
\left(\frac{1}{a_{\rm t}^3} \right)
\frac{3Gm_{\rm t} \sqrt{1 - e^2}}
{4 n \left(1 - e_{\rm t}^2\right)^{3/2}}
\nonumber
\\
&-&
\left(\frac{a}{a_{\rm t}^4} \right)
\frac{15Gm_{\rm t} e_{\rm t}\sqrt{1 - e^2} \left(4 + 9 e^2\right)}
{64 e n \left(1 - e_{\rm t}^2\right)^{5/2}}
\cos{\left(\varpi - \varpi_{\rm t}\right)}
\nonumber
\\
&+&
\left(\frac{a^2}{a_{\rm t}^5} \right)
\frac{45Gm_{\rm t} \sqrt{1 - e^2}}
{256 n \left(1 - e_{\rm t}^2\right)^{7/2}}
\big[
\left(4 + 3e^2\right)
\left(2 + 3e_{\rm t}^2\right)
\nonumber
\\
&+&
14
\left(1 + e^2\right)
e_{\rm t}^2
\cos{\left[2\left(\varpi - \varpi_{\rm t}\right)\right]}
\big]
+
\mathcal{O}\left(
\frac{Gm_{\rm t}}{n}
\frac{a^3}{a_{\rm t}^6} \right)
\label{doubavgp}
.
\end{eqnarray}

\noindent{}Equations (\ref{doubavga})-(\ref{doubavgp}) impart important information.

\begin{itemize}

\item{The leading order term $(1/a_{\rm t}^3)$ for the eccentricity variation vanishes, but that same term for the longitude of pericentre variation does not vanish.  Therefore, to leading order, the longitude of pericentre evolution of the secondary can be solved for exactly.  Consequently, ${\widetilde{\widehat{\varpi}}^{\rm P}}_{\rm sc} \propto t(1 - e_{\rm t})^{-3/2}$, illustrating that the precession rate is faster for highly eccentric tertiaries.  }
\item{The eccentricity variation appears to vanish at all orders when the tertiary's orbit is circular, whereas for the variation of the longitude of pericentre, only every other term vanishes in this limit.}
\item{The only term that is independent of both $\varpi$ and $\varpi_{\rm t}$ is the leading term for the variation of the longitude of pericentre.}
\item{Although all eccentricity variation terms appear to vanish when when $\varpi - \varpi_{\rm t} = 0$, this effect is instantaneous because the longitude of pericentre of the secondary is always precessing.}

\end{itemize}

\section{Averaging when tertiary is closer than secondary} \label{sfsec}

Now I consider the far secondary case, where the tertiary-primary
orbit is much smaller than the secondary-primary orbit.  
Assume both orbits are bounded orbits (osculating ellipses).  Here,
the primed semimajor axis always does vary after averaging, unlike
in the close secondary case.

\subsection{Averaging over tertiary orbit only} \label{sfsecsub1}

The leading order nonzero terms are 
long and I will not write them, but importantly I will indicate their 
leading nonzero order.  Note that the order of the expansion becomes
a function of the masses of the primary and tertiary.  Every other
of these terms vanishes in the special case of equal-mass binaries.
Also, unlike in Section \ref{sfsec}, in this subsection the evolution
of $f'$ must be averaged as well.

\subsubsection{Nonplanar equations}

\begin{eqnarray}
\widetilde{\left( \frac{da'}{dt} \right)}_{\rm sf} &=&
\mathcal{O}\left(
\frac{G}{n'}
\frac{a_{\rm t}^{2}}{r'^4}   \frac{m_{\rm t} m_{\rm p}}{ m_{\rm t}+m_{\rm p} }  \right)
,
\\
\widetilde{\left( \frac{de'}{dt} \right)}_{\rm sf} &=&
\mathcal{O}\left(
\frac{G}{n'}
\frac{a_{\rm t}^{2}}{a'^2r'^3} \frac{m_{\rm t} m_{\rm p}}{ m_{\rm t}+m_{\rm p} } \right)
,
\\
\widetilde{\left( \frac{di'}{dt} \right)}_{\rm sf} &=&
\mathcal{O}\left(
\frac{G}{n'}
\frac{a_{\rm t}^{2}}{a'^2r'^3} \frac{m_{\rm t} m_{\rm p}}{ m_{\rm t}+m_{\rm p} } \right)
,
\\
\widetilde{\left( \frac{d\Omega'}{dt} \right)}_{\rm sf} &=&
\mathcal{O}\left(
\frac{G}{n'}
\frac{a_{\rm t}^{2}}{a'^2r'^3} \frac{m_{\rm t} m_{\rm p}}{ m_{\rm t}+m_{\rm p} } \right)
,
\\
\widetilde{\left( \frac{d\omega'}{dt} \right)}_{\rm sf} &=&
\mathcal{O}\left(
\frac{G}{n'}
\frac{a_{\rm t}^{2}}{a'^2r'^3} \frac{m_{\rm t} m_{\rm p}}{ m_{\rm t}+m_{\rm p} } \right)
,
\end{eqnarray}

\begin{equation}
\widetilde{\left( \frac{df'}{dt} \right)}_{\rm sf} = 
\frac{n' \left(1 + e' \cos{f'}\right)^2}{\left(1 - e'^2\right)^{3/2}}
 - 
\widetilde{\left( \frac{d\omega'}{dt} \right)}_{\rm sf}
- \cos{i'} 
\widetilde{\left( \frac{d\Omega'}{dt} \right)}_{\rm sf}
.
\end{equation}

\subsubsection{Coplanar equations}

\begin{eqnarray}
{\widetilde{\left( \frac{da'}{dt} \right)}^{\rm P}}_{\rm sf} &=&
\mathcal{O}\left(
\frac{G}{n'}
\frac{a_{\rm t}^{2}}{r'^4}   \frac{m_{\rm t} m_{\rm p}}{ m_{\rm t}+m_{\rm p} }  \right)
,
\\
{\widetilde{\left( \frac{de'}{dt} \right)}^{\rm P}}_{\rm sf} &=&
\mathcal{O}\left(
\frac{G}{n'}
\frac{a_{\rm t}^{2}}{a'^2r'^3} \frac{m_{\rm t} m_{\rm p}}{ m_{\rm t}+m_{\rm p} } \right)
,
\\
{\widetilde{\left( \frac{d\varpi'}{dt} \right)}^{\rm P}}_{\rm sf} &=&
\mathcal{O}\left(
\frac{G}{n'}
\frac{a_{\rm t}^{2}}{a'^2r'^3} \frac{m_{\rm t} m_{\rm p}}{ m_{\rm t}+m_{\rm p} } \right)
,
\end{eqnarray}

\begin{equation}
{\widetilde{\left( \frac{df'}{dt} \right)}^{\rm P}}_{\rm sf} = 
\frac{n' \left(1 + e' \cos{f'}\right)^2}{\left(1 - e'^2\right)^{3/2}}
 - 
{\widetilde{\left( \frac{d\varpi'}{dt} \right)}^{\rm P}}_{\rm sf} 
.
\end{equation}

As shown, the change in semimajor axis is not zero, a marked difference from the previous averaged equations.

\subsection{Averaging over both orbits}

One may consider this case, at least to leading order, as a {\it reverse Lidov-Kozai} situation, where the perturber is internal.

\subsubsection{Nonplanar equations}

\begin{eqnarray}
\widehat{\widetilde{\left( \frac{da'}{dt} \right)}}_{\rm sf} 
&=& 
0
\label{LLa}
,
\\
\widehat{\widetilde{\left( \frac{de'}{dt} \right)}}_{\rm sf} 
&=& 
0 \times \left(\frac{a_{\rm t}^2}{a'^5} \frac{m_{\rm t} m_{\rm p}}{ m_{\rm t}+m_{\rm p} }\right)
\nonumber
\\
&+& 
\mathcal{O}\left(
\frac{G}{n'}
\frac{a_{\rm t}^3}{a'^6} \frac{m_{\rm t} m_{\rm p} \left( m_{\rm t} - m_{\rm p} \right)}{ \left(m_{\rm t}+m_{\rm p}\right)^2 }\right)
\label{LLe}
,
\\
\widehat{\widetilde{\left( \frac{di'}{dt} \right)}}_{\rm sf} 
&=&
\left( \frac{a_{\rm t}^2}{a'^5} \frac{m_{\rm t} m_{\rm p}}{m_{\rm t}+m_{\rm p}}\right)
\frac{15 G e_{\rm t}^2 \sin{i'} \sin{\left(2\Omega' - 2\varpi_{\rm t} \right)} }
{8 n' \left(1 - e'^2 \right)^{2}}
\nonumber
\\
&+&
\mathcal{O}\left(
\frac{G}{n'}
\frac{a_{\rm t}^3}{a'^6} \frac{m_{\rm t} m_{\rm p} \left( m_{\rm t} - m_{\rm p} \right)}{ \left(m_{\rm t}+m_{\rm p}\right)^2 }\right)
\label{LLi}
,
\\
\widehat{\widetilde{\left( \frac{d\Omega'}{dt} \right)}}_{\rm sf} 
&=& 
\left( \frac{a_{\rm t}^2}{a'^5} \frac{m_{\rm t} m_{\rm p}}{m_{\rm t}+m_{\rm p}}\right)
\frac{3 G \cos{i'} \left(-2 - 3e_{\rm t}^2  + 5e_{\rm t}^2 \cos{\left(2\Omega' - 2\varpi_{\rm t} \right)}  \right) }
{8 n' \left(1 - e'^2 \right)^{2}}
\nonumber
\\
&+&
\mathcal{O}\left(
\frac{G}{n'}
\frac{a_{\rm t}^3}{a'^6} \frac{m_{\rm t} m_{\rm p} \left( m_{\rm t} - m_{\rm p} \right)}{ \left(m_{\rm t}+m_{\rm p}\right)^2 }\right)
\label{LLOm}
,
\\
\widehat{\widetilde{\left( \frac{d\omega'}{dt} \right)}}_{\rm sf} 
&=& 
-\left( \frac{a_{\rm t}^2}{a'^5} \frac{m_{\rm t} m_{\rm p}}{m_{\rm t}+m_{\rm p}}\right)
\left(\frac{3 G}{32 n' \left(1 - e'^2 \right)^{2}}\right)
\nonumber
\\
&\times&
\bigg\lbrace
5 e_{\rm t}^2 \left(5 \cos{\left(2i'\right)} - 1 \right)
\cos{\left[\left(2\Omega' - 2\varpi_{\rm t} \right)\right]}
-
\left(2 + 3 e_{\rm t}^2 \right)
\left(3 + 5 \cos{\left(2i'\right)}  \right)
\bigg\rbrace
\nonumber
\\
&+&
\mathcal{O}\left(
\frac{G}{n'}
\frac{a_{\rm t}^3}{a'^6} \frac{m_{\rm t} m_{\rm p} \left( m_{\rm t} - m_{\rm p} \right)}{ \left(m_{\rm t}+m_{\rm p}\right)^2 }\right)
\label{LLom}
\end{eqnarray}

The leading order term for the semimajor axis and eccentricity evolution vanishes (equations \ref{LLa}-\ref{LLe}), but not for the evolution of the inclination, longitude of ascending node, nor the argument of pericentre.  This striking observation allows me to consider obtaining a complete solution or stationary solution to these equations to leading order.  Such solutions would also hold if the next order is included for equal-mass binaries, which would cause those terms to vanish.  

Also striking is that to leading order, none of the equations are dependent on $\omega'$.  Therefore, the problem reduces to two variables and two equations (equations \ref{LLi}-\ref{LLOm}).  I cannot find a complete solution, but at least one stationary solution does exist, when the secondary is on a polar orbit and the longitude of ascending node or longitude of descending node is equal to $\varpi_{\rm t}$.  When this configuration occurs, the argument of pericentre will still precess while the inclination and longitude of ascending node will remain static.  The argument of pericentre will also become stationary at the critical value $e_{\rm t,crit} = \sqrt{1/6} \approx 0.41$.

Equations (\ref{LLe})-(\ref{LLom}) are in fact similar to their unprimed Lidov-Kozai counterparts (equations \ref{Koze}-\ref{Kozw}).  The largest difference is that the variation in $e'$ vanishes to leading order.  The other variables have similar dependencies and forms except that notably all the leading-order primed variable terms are dependent on the secondary's longitude of ascending node.

\subsubsection{Coplanar equations}

\begin{eqnarray}
{\widehat{\widetilde{\left( \frac{da'}{dt} \right)}}^{\rm P}}_{\rm sf} 
&=& 
0
,
\\
{\widehat{\widetilde{\left( \frac{de'}{dt} \right)}}^{\rm P}}_{\rm sf} 
&=& 
0 \times
\left(\frac{a_{\rm t}^2}{a'^5} \right) 
\left(\frac{m_{\rm t} m_{\rm p}}{m_{\rm t} + m_{\rm p}}\right)
\nonumber
\\
&-&
\left( \frac{a_{\rm t}^3}{a'^6} \frac{m_{\rm t} m_{\rm p} \left(m_{\rm p} - m_{\rm t} \right)}{\left(m_{\rm t}+m_{\rm p}\right)^2}\right)
\frac{15 G e_{\rm t} \left(4 + 3 e_{\rm t}^2\right) \sin{\left(\varpi' - \varpi_{\rm t} \right)} }
{64 n \left(1 - e'^2 \right)^{2}}
\nonumber
\\
&+&
\mathcal{O}\left(
\frac{G}{n'}
\frac{a_{\rm t}^4}{a'^7}  \frac{m_{\rm t} m_{\rm p}\left(m_{\rm p}^2 + m_{\rm t}^2 -m_{\rm p} m_{\rm t} \right)}{\left(m_{\rm t} + m_{\rm p}\right)^3}  \right)
,
\\
{\widehat{\widetilde{\left( \frac{d\varpi'}{dt} \right)}}^{\rm P}}_{\rm sf} 
&=& 
-
\left(\frac{a_{\rm t}^2}{a'^5} \right) 
\left(\frac{m_{\rm t} m_{\rm p}}{m_{\rm t} + m_{\rm p}}\right)
\left[
\frac{3 G \left(2 + 3e_{\rm t}^2\right)}{8 n' \left(1-e'^2\right)^2}
\right]
\nonumber
\\
&-&
\left(\frac{a_{\rm t}^3}{a'^6} \right) 
\left( \frac{m_{\rm t} m_{\rm p}\left(m_{\rm t} - m_{\rm p}\right)}{\left(m_{\rm t} + m_{\rm p}\right)^2}  \right)
\nonumber
\\
&\times&
\left[
\frac{15 G e_{\rm t} \left(4 + 3e_{\rm t}^2\right)\left(1 + 4e'^2\right) \cos{\left(\varpi - \varpi_{\rm t} \right)}}{64 n' e'\left(1-e'^2\right)^3}
\right]
\nonumber
\\
&+&
\mathcal{O}
\left(
\frac{G}{n'}
\frac{a_{\rm t}^4}{a'^7}  \frac{m_{\rm t} m_{\rm p}\left(m_{\rm p}^2 + m_{\rm t}^2 -m_{\rm p} m_{\rm t} \right)}{\left(m_{\rm t} + m_{\rm p}\right)^3}  \right)
\label{exact2}
\end{eqnarray}

As in the unprimed case, the evolution of the argument of pericentre dominates the change in the secondary's orbit.  The change in eccentricity is a higher-order effect.

\section{The Circular Circumbinary Case}\label{circum}

Stellar systems containing tight binaries are common.  These binaries usually have tidally circularized, which means that their current orbit is a circle.  The behaviour of any particles or planets external to the binary is described by the primed element evolution equations in the special case of $e_{\rm t} = 0$.  This section presents this special case because of its importance and because the resulting simplification to the equations is significant, providing new insights into the motion.

\subsection{Averaging over the binary companion (tertiary) only}

Unlike in subsection \ref{sfsecsub1}, here the leading-order term of each equation is compact enough to present and potentially be useful.

\subsubsection{Nonplanar equations}

\begin{eqnarray}
\widetilde{\left( \frac{da'}{dt} \right)}_{\rm sf}\bigg|_{e_{\rm t} = 0} &=&
-
\left(\frac{a_{\rm t}^{2}}{r'^4} \right)
\left( \frac{m_{\rm t} m_{\rm p}}{ m_{\rm t}+m_{\rm p} }  \right)
\left(\frac{3G}{2n' \left(1-e'^2\right)^{1/2}}\right)
\nonumber
\\
&\times&
\bigg\lbrace
\big[e' \sin{f'} \left(5 D_{{\rm t},3}^{'2}+5 D_{{\rm t},4}^{'2}-2\right)
\nonumber
\\
&+&
2 D_{{\rm t},3}' D_{{\rm t},5}'-2 D_{{\rm t},4}' D_{{\rm t},6}'\big]
\bigg\rbrace 
\nonumber
\\
&+& 
\mathcal{O}\left(
\frac{G}{n'}
\frac{a_{\rm t}^{3}}{r'^5}     \frac{m_{\rm t} m_{\rm p}\left(m_{\rm t} - m_{\rm p}\right)}{\left(m_{\rm t} + m_{\rm p}\right)^2}   \right)
,
\label{circumgena}
\\
\widetilde{\left( \frac{de'}{dt} \right)}_{\rm sf}\bigg|_{e_{\rm t} = 0} &=&
\left(\frac{a_{\rm t}^{2}}{a'^2r'^3} \right)
\left( \frac{m_{\rm t} m_{\rm p}}{ m_{\rm t}+m_{\rm p} }  \right)
\left(\frac{3G}{8n' \left(1-e'^2\right)^{1/2}}\right)
\nonumber
\\
&\times&
\bigg\lbrace
\sin{f'} \left(-5 D_{{\rm t},3}^{'2}-5 D_{{\rm t},4}^{'2}+2\right) (e' \cos{f'}+1)
\nonumber
\\
&-&
D_{{\rm t},3}' D_{{\rm t},7}'+D_{{\rm t},4}' D_{{\rm t},8}'+ e'\sin{\left(2f'\right)} 
\bigg\rbrace
\nonumber
\\
&+&
\mathcal{O}
\left(
\frac{G}{n'}
\frac{a_{\rm t}^{3}}{a'^2r'^4}  \frac{m_{\rm t} m_{\rm p}\left(m_{\rm t} - m_{\rm p}\right)}{\left(m_{\rm t} + m_{\rm p}\right)^2} \right)
,
\\
\widetilde{\left( \frac{di'}{dt} \right)}_{\rm sf}\bigg|_{e_{\rm t} = 0} &=&
-
\left(\frac{a_{\rm t}^{2}}{a'^2r'^3} \right)
\left( \frac{m_{\rm t} m_{\rm p}}{ m_{\rm t}+m_{\rm p} }  \right)
\left(\frac{3G}{2n' \left(1-e'^2\right)^{1/2}}\right)
\nonumber
\\
&\times&
\sin{i'} \cos (f'+\omega' ) 
\nonumber
\\
&\times&
\bigg\lbrace
D_{t,4}' \cos{\Omega'} -D_{t,3}' \sin{\Omega'}
\bigg\rbrace
\nonumber
\\
&+&
\mathcal{O}
\left(
\frac{G}{n'}
\frac{a_{\rm t}^{3}}{a'^2r'^4}\frac{m_{\rm t} m_{\rm p}\left(m_{\rm t} - m_{\rm p}\right)}{\left(m_{\rm t} + m_{\rm p}\right)^2}  \right)
,
\\
\widetilde{\left( \frac{d\Omega'}{dt} \right)}_{\rm sf}\bigg|_{e_{\rm t} = 0} &=&
-
\left(\frac{a_{\rm t}^{2}}{a'^2r'^3} \right)
\left( \frac{m_{\rm t} m_{\rm p}}{ m_{\rm t}+m_{\rm p} }  \right)
\left(\frac{3G}{2n' \left(1-e'^2\right)^{1/2}}\right)
\nonumber
\\
&\times&
\sin (f'+\omega' )
\nonumber
\\
&\times&
\bigg\lbrace
D_{t,4}' \cos{\Omega'} -D_{t,3}' \sin{\Omega'}
\bigg\rbrace
\nonumber
\\
&+&
\mathcal{O}
\left(
\frac{G}{n'}
\frac{a_{\rm t}^{3}}{a'^2r'^4} \frac{m_{\rm t} m_{\rm p}\left(m_{\rm t} - m_{\rm p}\right)}{\left(m_{\rm t} + m_{\rm p}\right)^2} \right)
,
\\
\widetilde{\left( \frac{d\omega'}{dt} \right)}_{\rm sf}\bigg|_{e_{\rm t} = 0} &=&
\left(\frac{a_{\rm t}^{2}}{a'^2r'^3} \right)
\left( \frac{m_{\rm t} m_{\rm p}}{ m_{\rm t}+m_{\rm p} }  \right)
\left(\frac{3G}{8 e' n' \left(1-e'^2\right)^{1/2}}\right)
\nonumber
\\
&\times&
\bigg\lbrace
\left(5 D_{t,3}^{'2}+5 D_{t,4}^{'2}-2\right) (e' \cos (2 f')+2 \cos {f'})
\nonumber
\\
&+&
5 e' D_{t,3}^{'2}+2 D_{t,3}' D_{t,9}'-2 D_{t,4}' D_{t,10}'-2 e'  
\bigg\rbrace
\nonumber
\\
&+&
\mathcal{O}
\left(
\frac{G}{n'}
\frac{a_{\rm t}^{3}}{a'^2r'^4} \frac{m_{\rm t} m_{\rm p}\left(m_{\rm t} - m_{\rm p}\right)}{\left(m_{\rm t} + m_{\rm p}\right)^2} \right)
,
\end{eqnarray}

\begin{equation}
\widetilde{\left( \frac{df'}{dt} \right)}_{\rm sf}\bigg|_{e_{\rm t} = 0} = 
\frac{n' \left(1 + e' \cos{f'}\right)^2}{\left(1 - e'^2\right)^{3/2}}
 - 
\widetilde{\left( \frac{d\omega'}{dt} \right)}_{\rm sf}\bigg|_{e_{\rm t} = 0}
- \cos{i'} 
\widetilde{\left( \frac{d\Omega'}{dt} \right)}_{\rm sf}\bigg|_{e_{\rm t} = 0}
.
\label{circumgenf}
\end{equation}

In equations (\ref{circumgena})-(\ref{circumgenf}), the value of $\varpi_{\rm t}$ in
the auxiliary $D$ variables should be set to zero.  The reason goes back to equations
(\ref{xjGEN})-(\ref{zjGEN}); if the tertiary has a forever circular orbit, $x_{\rm t}$ 
and $y_{\rm t}$ are parameterized by $f_{\rm t}$ alone.  In this case, 
$f_{\rm t} = \Pi_{\rm t} = n_{\rm t} t$, where $\Pi_{\rm t}$ represents the tertiary's
mean anomaly.

\subsubsection{Coplanar equations}

\begin{eqnarray}
{\widetilde{\left( \frac{da'}{dt} \right)}^{\rm P}}_{\rm sf}\bigg|_{e_{\rm t} = 0} &=&
-
\left(\frac{a_{\rm t}^{2}}{r'^4} \right)
\left( \frac{m_{\rm t} m_{\rm p}}{ m_{\rm t}+m_{\rm p} }  \right)
\left(\frac{3G}{2n' \left(1-e'^2\right)^{1/2}}\right)
\nonumber
\\
&\times&
\bigg\lbrace
\big[e' \sin{f'} \left(5  \left[D_{{\rm t},3}^{'\rm P}\right]^2 +5 \left[D_{{\rm t},4}^{'\rm P}\right]^2  -2\right)
\nonumber
\\
&+&
2 D_{{\rm t},3}^{'\rm P} D_{{\rm t},5}^{'\rm P}-2 D_{{\rm t},4}^{'\rm P} D_{{\rm t},6}^{'\rm P}\big]
\bigg\rbrace 
\nonumber
\\
&+& 
\mathcal{O}\left(
\frac{G}{n'}
\frac{a_{\rm t}^{3}}{r'^5}     \frac{m_{\rm t} m_{\rm p}\left(m_{\rm t} - m_{\rm p}\right)}{\left(m_{\rm t} + m_{\rm p}\right)^2}   \right)
,
\\
{\widetilde{\left( \frac{de'}{dt} \right)}^{\rm P}}_{\rm sf}\bigg|_{e_{\rm t} = 0} &=&
\left(\frac{a_{\rm t}^{2}}{a'^2r'^3} \right)
\left( \frac{m_{\rm t} m_{\rm p}}{ m_{\rm t}+m_{\rm p} }  \right)
\left(\frac{3G}{8n' \left(1-e'^2\right)^{1/2}}\right)
\nonumber
\\
&\times&
\bigg\lbrace
\sin{f'} \left(-5  \left[D_{{\rm t},3}^{'\rm P}\right]^2 -5 \left[D_{{\rm t},4}^{'\rm P}\right]^2 +2\right) (e' \cos{f'}+1)
\nonumber
\\
&-&
D_{{\rm t},3}^{'\rm P} D_{{\rm t},7}^{'\rm P}+D_{{\rm t},4}^{'\rm P} D_{{\rm t},8}^{'\rm P}+ e'\sin{\left(2f'\right)} 
\bigg\rbrace
\nonumber
\\
&+&
\mathcal{O}
\left(
\frac{G}{n'}
\frac{a_{\rm t}^{3}}{a'^2r'^4}  \frac{m_{\rm t} m_{\rm p}\left(m_{\rm t} - m_{\rm p}\right)}{\left(m_{\rm t} + m_{\rm p}\right)^2} \right)
,
\\
{\widetilde{\left( \frac{d\varpi'}{dt} \right)}^{\rm P}}_{\rm sf}\bigg|_{e_{\rm t} = 0} &=&
-
\left(\frac{a_{\rm t}^{2}}{a'^2r'^3} \right)
\left( \frac{m_{\rm t} m_{\rm p}}{ m_{\rm t}+m_{\rm p} }  \right)
\left(\frac{3 G}{8 n' e' \left(1-e'^2\right)^{1/2}   } \right)
\nonumber
\\
&\times&
\bigg\lbrace
-2 D_{{\rm t},11}' D_{{\rm t},4}^{'\rm P}+2 D_{{\rm t},12}' D_{{\rm t},3}^{'\rm P}
\nonumber
\\
&+&
\left(-5 \left[D_{{\rm t},3}^{'\rm P}\right]^2-5 \left[D_{{\rm t},4}^{'\rm P}\right]^2+2\right) (e' \cos (2 f')+2 \cos {f'})
\nonumber
\\
&-&
5 e'  \left( \left[D_{{\rm t},3}^{'\rm P}\right]^2    +   \left[D_{{\rm t},4}^{'\rm P}\right]^2   \right)+2 e'  
\bigg\rbrace
\nonumber
\\
&+&
\mathcal{O}
\left(
\frac{G}{n'}
\frac{a_{\rm t}^{3}}{a'^2r'^4} \frac{m_{\rm t} m_{\rm p}\left(m_{\rm t} - m_{\rm p}\right)}{\left(m_{\rm t} + m_{\rm p}\right)^2} \right)
\end{eqnarray}

\begin{equation}
{\widetilde{\left( \frac{df'}{dt} \right)}^{\rm P}}_{\rm sf}\bigg|_{e_{\rm t} = 0} = 
\frac{n' \left(1 + e' \cos{f'}\right)^2}{\left(1 - e'^2\right)^{3/2}}
 - 
{\widetilde{\left( \frac{d\varpi'}{dt} \right)}^{\rm P}}_{\rm sf}\bigg|_{e_{\rm t} \rightarrow 0}
\end{equation}

\subsection{Averaging over both orbits}

The following equations highlight the importance of the higher-order functional dependencies of the binary masses.  For some perspective about how steeply the magnitude of the mass function changes, successive orders for equal-mass binaries yield $\left(\frac{m_{\rm p}}{2}, 0, \frac{m_{\rm p}}{8}, 0, \frac{m_{\rm p}}{32} \right)$ and for $m_{\rm p} = 2 m_{\rm t}$ yield 

\noindent{}$\left(\frac{m_{\rm p}}{3}, \frac{m_{\rm p}}{9}, \frac{m_{\rm p}}{9}, \frac{m_{\rm p}}{27}, \frac{11m_{\rm p}}{273} \right)$.

\subsubsection{Nonplanar equations}

\begin{eqnarray}
\widehat{\widetilde{\left( \frac{da'}{dt} \right)}}_{\rm sf}\bigg|_{e_{\rm t} = 0}
&=& 
0 
,
\\
\widehat{\widetilde{\left( \frac{de'}{dt} \right)}}_{\rm sf}\bigg|_{e_{\rm t} = 0}
&=& 
0 \times \left(\frac{a_{\rm t}^2}{a'^5} \right) 
\left(\frac{m_{\rm t} m_{\rm p}}{m_{\rm t} + m_{\rm p}} \right)
\nonumber
\\
&+&
0 \times \left(\frac{a_{\rm t}^3}{a'^6} \right) 
\left( \frac{m_{\rm t} m_{\rm p}\left(m_{\rm t} - m_{\rm p}\right)}{\left(m_{\rm t} + m_{\rm p}\right)^2} \right)
\nonumber
\\
&+&
\left(\frac{a_{\rm t}^4}{a'^7} \right) 
\left(\frac{m_{\rm t} m_{\rm p}\left(m_{\rm p}^2 + m_{\rm t}^2 - m_{\rm p} m_{\rm t}\right)}{\left(m_{\rm t} + m_{\rm p}\right)^3}\right)
\nonumber
\\
&\times&
\frac{45 G e' \sin^2{i'} \sin{\left(2\omega'\right)} \left(5 + 7 \cos{\left(2 i'\right)} \right)}
{512 n' \left(1 - e'^2 \right)^3}
\nonumber
\\
&+& \mathcal{O}
\left(\frac{G}{n'}
\frac{a_{\rm t}^5}{a'^8} 
\frac{m_{\rm t} m_{\rm p}\left(m_{\rm p}^3 + m_{\rm t}^3 - m_{\rm p}^2 m_{\rm t} - m_{\rm p} m_{\rm t}^2\right)}{\left(m_{\rm t} + m_{\rm p}\right)^4}\right)
,
\label{gene2avgcircum}
\\
\widehat{\widetilde{\left( \frac{di'}{dt} \right)}}_{\rm sf}\bigg|_{e_{\rm t} \rightarrow 0}
&=& 
0 \times \left(\frac{a_{\rm t}^2}{a'^5} \right) 
\left(\frac{m_{\rm t} m_{\rm p}}{m_{\rm t} + m_{\rm p}} \right)
\nonumber
\\
&+&
0 \times \left(\frac{a_{\rm t}^3}{a'^6} \right) 
\left( \frac{m_{\rm t} m_{\rm p}\left(m_{\rm t} - m_{\rm p}\right)}{\left(m_{\rm t} + m_{\rm p}\right)^2} \right)
\nonumber
\\
&-&
\left(\frac{a_{\rm t}^4}{a'^7} \right) 
\left(\frac{m_{\rm t} m_{\rm p}\left(m_{\rm p}^2 + m_{\rm t}^2 - m_{\rm p} m_{\rm t}\right)}{\left(m_{\rm t} + m_{\rm p}\right)^3}\right)
\nonumber
\\
&\times&
\frac{45 G e'^2 \sin{2i'} \sin{\left(2\omega'\right)} \left(5 + 7 \cos{\left(2 i'\right)} \right)}
{1024 n' \left(1 - e'^2 \right)^4}
\nonumber
\\
&+& \mathcal{O}
\left(
\frac{G}{n'}
\frac{a_{\rm t}^5}{a'^8} 
\frac{m_{\rm t} m_{\rm p}\left(m_{\rm p}^3 + m_{\rm t}^3 - m_{\rm p}^2 m_{\rm t} - m_{\rm p} m_{\rm t}^2\right)}{\left(m_{\rm t} + m_{\rm p}\right)^4}\right)
,
\\
\widehat{\widetilde{\left( \frac{d\Omega'}{dt} \right)}}_{\rm sf}\bigg|_{e_{\rm t} = 0}
&=& 
-
\left(\frac{a_{\rm t}^2}{a'^5} \right) 
\left(\frac{m_{\rm t} m_{\rm p}}{m_{\rm t} + m_{\rm p}} \right)
\frac{3 G \cos{i'}}
{4 n' \left(1 - e'^2 \right)^2}
\nonumber
\\
&+&
0 \times \left(\frac{a_{\rm t}^3}{a'^6} \right) 
\left( \frac{m_{\rm t} m_{\rm p}\left(m_{\rm t} - m_{\rm p}\right)}{\left(m_{\rm t} + m_{\rm p}\right)^2} \right)
\nonumber
\\
&+&
\left(\frac{a_{\rm t}^4}{a'^7} \right) 
\left(\frac{m_{\rm t} m_{\rm p}\left(m_{\rm p}^2 + m_{\rm t}^2 - m_{\rm p} m_{\rm t}\right)}{\left(m_{\rm t} + m_{\rm p}\right)^3}\right)
\nonumber
\\
&\times&
\frac{G}
{1024 n' \left(1 - e'^2 \right)^4}
\bigg[
90 e'^2 \cos{\left(2\omega'\right)} \left(5 \cos{i'} + 7 \cos{\left(3i'\right)} \right)
\nonumber
\\
&-&
45 \left(2 + 3e'^2\right) \left(9 \cos{i'} + 7 \cos{\left(3i'\right)}  \right)
\bigg]
\nonumber
\\
&+& \mathcal{O}
\left(
\frac{G}{n'}
\frac{a_{\rm t}^5}{a'^8} 
\frac{m_{\rm t} m_{\rm p}\left(m_{\rm p}^3 + m_{\rm t}^3 - m_{\rm p}^2 m_{\rm t} - m_{\rm p} m_{\rm t}^2\right)}{\left(m_{\rm t} + m_{\rm p}\right)^4}\right)
,
\label{genOm2avgcircum}
\\
\widehat{\widetilde{\left( \frac{d\omega'}{dt} \right)}}_{\rm sf}\bigg|_{e_{\rm t} = 0}
&=& 
\left(\frac{a_{\rm t}^2}{a'^5} \right) 
\left(\frac{m_{\rm t} m_{\rm p}}{m_{\rm t} + m_{\rm p}} \right)
\frac{3 G \left(3 + 5 \cos{\left(2i'\right)} \right)}
{16 n' \left(1 - e'^2 \right)^2}
\nonumber
\\
&+&
0 \times \left(\frac{a_{\rm t}^3}{a'^6} \right) 
\left( \frac{m_{\rm t} m_{\rm p}\left(m_{\rm t} - m_{\rm p}\right)}{\left(m_{\rm t} + m_{\rm p}\right)^2} \right)
\nonumber
\\
&+&
\left(\frac{a_{\rm t}^4}{a'^7} \right) 
\left(\frac{m_{\rm t} m_{\rm p}\left(m_{\rm p}^2 + m_{\rm t}^2 - m_{\rm p} m_{\rm t}\right)}{\left(m_{\rm t} + m_{\rm p}\right)^3}\right)
\nonumber
\\
&\times&
\frac{45G}
{8129 n' \left(1 - e'^2 \right)^4}
\nonumber
\\
&\times&
\bigg[
4 \cos{2i'} \left(52 + 63e'^2 + \cos{\left(2\omega'\right)} \left(4 - 14e'^2 \right)  \right)
\nonumber
\\
&-&
7 \cos{4i'} 
\left[
2 \cos{\left(2\omega'\right)} \left(2 + 9 e'^2\right) - 27 e'^2 - 28
\right]
\nonumber
\\
&+&
2\cos{\left(2 \omega'\right)}
\left(6 - 5e'^2\right)
+
27 \left(4 + 5e'^2\right)
\bigg]
\nonumber
\\
&+& \mathcal{O}
\left(\frac{G}{n'}
      \frac{a_{\rm t}^5}{a'^8} 
\frac{m_{\rm t} m_{\rm p}\left(m_{\rm p}^3 + m_{\rm t}^3 - m_{\rm p}^2 m_{\rm t} - m_{\rm p} m_{\rm t}^2\right)}{\left(m_{\rm t} + m_{\rm p}\right)^4}\right)
.
\label{genom2avgcircum}
\end{eqnarray}

Notably, the leading-order nonzero terms for the change in $\Omega'$ and $\omega'$ are $\left(a_{\rm t}^2/a'^5\right)$, whereas for $e'$ and $i'$ the terms are of order $\left(a_{\rm t}^4/a'^7\right)$.  Hence, the orbital change is dominated by orientation variations as opposed to stretching or warping.  The two leading order terms are immediately solvable because neither are functions of $\Omega'$ nor $\omega'$.  Hence, the precession of both the pericentre and node proceed linearly with time to an excellent approximation.  The precession rate is dependent on $a'$, $e'$ and $i'$.  This rate vanishes for $\Omega'$ only for polar orbits, and for $\omega'$ only at a critical value $i'_{\rm crit} = (1/2) \cos^{-1}{\left[-3/5\right]} \approx 63.4^{\circ}$.

\subsubsection{Coplanar equations}

The nonzero leading-order term in equation (\ref{gene2avgcircum}) vanishes when $i'=0$, suggesting that the planar equations in circular circumbinary systems afford even greater simplification.

\begin{eqnarray}
{\widehat{\widetilde{\left( \frac{da'}{dt} \right)}}^{\rm P}}_{\rm sf}\bigg|_{e_{\rm t} = 0} 
&=&
0
\\
{\widehat{\widetilde{\left( \frac{de'}{dt} \right)}}^{\rm P}}_{\rm sf}\bigg|_{e_{\rm t} = 0} 
&=&
0
\label{solveecc}
\\
{\widehat{\widetilde{\left( \varpi' \right)}}^{\rm P}}_{\rm sf}\bigg|_{e_{\rm t} = 0} 
&=& 
t \left(\frac{a_{\rm t}^2}{a'^5} \right) 
\left(\frac{m_{\rm t} m_{\rm p}}{m_{\rm t} + m_{\rm p}}\right)
\left[
\frac{3 G}
{4 n' \left(1 - e'^2\right)^2}
\right]
\nonumber
\\
&+&
0 \times \left(\frac{a_{\rm t}^3}{a'^6} \right) 
\left( \frac{m_{\rm t} m_{\rm p}\left(m_{\rm t} - m_{\rm p}\right)}{\left(m_{\rm t} + m_{\rm p}\right)^2} \right)
\nonumber
\\
&+&
t
\left(\frac{a_{\rm t}^4}{a'^7} \right)
\left[\frac{m_{\rm t} m_{\rm p}\left(m_{\rm p}^2 + m_{\rm t}^2 - m_{\rm p} m_{\rm t}\right)}{\left(m_{\rm t} + m_{\rm p}\right)^3}\right]
\nonumber
\\
&\times&
\left[ 
\frac{45 G \left(4 + 3e'^2\right)}
{128 n' \left(1 - e'^2\right)^4}
\right]
\nonumber
\\
&+&
0 \times \left(\frac{a_{\rm t}^5}{a'^8} \right) 
\left( \frac{m_{\rm t} m_{\rm p} \left( m_{\rm p}^3 + m_{\rm t}^3 - m_{\rm p}^2m_{\rm t} - m_{\rm p}m_{\rm t}^2 \right)  }{\left(m_{\rm t} + m_{\rm p}\right)^4} \right)
\nonumber
\\
&+&
t
\left(\frac{a_{\rm t}^6}{a'^9} \right)
\left[\frac{m_{\rm t} m_{\rm p}\left( m_{\rm p}^4 + m_{\rm t}^4 - m_{\rm p}^3m_{\rm t} - m_{\rm t}^3m_{\rm p} + m_{\rm t}^2m_{\rm p}^2\right)}
{\left(m_{\rm t} + m_{\rm p}\right)^5}\right]
\nonumber
\\
&\times&
\left[
\frac{525 G \left[8 + 5e'^2 \left(4 + e'^2 \right) \right]}
{2048 n' \left(1 - e'^2\right)^6}
\right]
+
...
\label{solveexact}
\end{eqnarray}

Equation (\ref{solveexact}) is not written as a differential equation because a full solution is available out to an order of at least 
$\left(a_{\rm t}^6/a'^9\right)$.  This solution, which is linear with time, exists because the evolution of the argument of
pericentre is independent of itself.  The dependence, which exists in the nonplanar version (equation \ref{genom2avgcircum}), vanishes in the coplanar limit.  Hence, I achieve a complete solution to the first three nonzero orders.  Computing additional terms becomes challenging, and may not be particularly useful.  What is useful is that equation (\ref{solveexact}) satisfies an arbitrarily high value of $e'$.

Nevertheless, the equation provides a good opportunity to link this formalism to established dynamical theory.
Now I demonstrate how equations (\ref{solveecc}-\ref{solveexact}) reduce to coplanar Laplace-Lagrange secular theory, a popular treatment of which is described in Chapter 7 of \cite{murder1999}.  The theory enables one to obtain approximate doubly-averaged solutions of the equations of motion in the three-body problem in limited situations.  One limitation is that the eccentricities of the bodies must be small.  The classic treatment expands the eccentricities out to second order.  In contrast, equation (\ref{solveexact}) satisfies any value of $e'$.

Rather than compare the results to the classic theory, I compare the results to the more expansive fourth-order Laplace-Lagrange theory (Veras \& Armitage 2007).  Equation (8) of that paper provides the necessary disturbing function, to be used in conjunction with the Lagrange's equations of motion in their equations (10c) and (10d), which importantly differ from the standard reduced versions given in equations (7.16) of \cite{murder1999}.  Retaining disturbing function eccentricity terms to fourth order and expanding equations (10c) and (10d) from \cite{verarm2007} about small eccentricity yields exactly the same coefficients from equation (\ref{solveexact}) of this paper when expanded about small eccentricity, to at least the first two nonzero orders in semimajor axis ratio and eccentricity. The mass functions naturally differ because of the different setups. Also, note that the secondary eccentricity evolution similarly vanishes in coplanar Laplace-Lagrange theory when $e_{\rm t} = 0$.

\section{Resonances} \label{ress}

Mean motion resonances may occur in the restricted three-body problem just as in the
unrestricted problem.  Notably, none of the equations presented so far require a 
distinction to be made; the equations satisfy both resonant and non-resonant behaviour.
Now I place the formalism in the context of resonances.

In a three body system, a mean motion resonance between the secondary and tertiary
is helped defined by the following time-dependent angle 

\begin{equation}
\phi = q_{_{\lambda_{\rm out}}}\lambda_{\rm out} + q_{_{\lambda_{\rm in}}}\lambda_{\rm in} 
     + q_{_{\lambda_{\rm out}}}\varpi_{\rm out} + q_{_{\lambda_{\rm in}}}\varpi_{\rm in}
     + q_{_{\lambda_{\rm out}}}\Omega_{\rm out} + q_{_{\lambda_{\rm in}}}\Omega_{\rm in}
\end{equation}

\noindent{}where the mean longitude $\lambda \equiv \varpi + \Pi$,
the mean anomaly $\Pi \equiv E - e \sin{E}$, $E$ is the eccentric anomaly,  
and the constants $q$ add to zero.  The subscripts ``in'' and ``out'' refer
to the chosen ordering of the secondary and tertiary with respect to distance
from the primary.

The time evolution of $\phi$ determines whether the secondary and tertiary
are in a particular single-argument resonance defined by the $q$ values.  Typical analytical
treatments of obtaining $\dot{\phi}$ utilize a disturbing function that is 
truncated in orders of eccentricity, where a dot denotes a time derivative. Here, I show 
how an explicit relation for $\dot{\phi}$ with arbitrarily high eccentricities is obtained 
using the equations in this paper. 

If the (zero-mass) secondary is the inner body, then its orbital elements are measured
with respect to the primary.  If the secondary is the outer body, then instead the
primed elements should be used, which are measured with respect to the centre of 
mass of the primary and tertiary.  Without loss of generality for this exercise, 
assume the former case.  The tertiary's orbit will never change, and hence 
$\dot{\varpi}_{\rm out} = \dot{\Omega}_{\rm out} = 0$.  Also, 
$\dot{\lambda}_{\rm out} = \dot{\Pi}_{\rm out} = n_{\rm out}$ is known.

For the secondary, I obtain

\begin{equation}
\dot{\lambda}_{\rm in} = \dot{\omega}_{\rm in} + \dot{\Omega}_{\rm in}
+ \dot{E}_{\rm in} \left(1 - e_{\rm in} \cos{E_{\rm in}}\right) 
- \dot{e} \sin{E_{\rm in}}
.
\end{equation}

\noindent{}Also, from equation (12) of \cite{vereva2013}

\begin{equation}
\dot{E}_{\rm in} = 
\frac{\sqrt{1 - e_{\rm in}^2}}{1 + e_{\rm in}\cos{f_{\rm in}} } \dot{f}_{\rm in}
 - 
\frac{\sin{f_{\rm in}}}{\sqrt{1 - e_{\rm in}^2} \left(1 + e_{\rm in}\cos{f_{\rm in}}  \right)  }
\dot{e}_{\rm in}
.
\label{deindf}
\end{equation}

Using equation (\ref{deindf}) along with the standard relations

\begin{equation}
\cos{E_{\rm in}} = \frac{e_{\rm in} + \cos{f_{\rm in}}}
{1 + e_{\rm in} \cos{f_{\rm in}} },
   \   \   \  \  \
\sin{E_{\rm in}} = 
\frac{\sin{f_{\rm in}}\sqrt{1 - e_{\rm in}^2}}
{1 + e_{\rm in} \cos{f_{\rm in}}  }
\end{equation}

\noindent{gives}

\begin{equation}
\dot{\lambda}_{\rm in} = \dot{\omega}_{\rm in} + \dot{\Omega}_{\rm in}
+   \frac{\left(1 - e_{\rm in}^2\right)^{3/2}}{\left(1 + e_{\rm in} \cos{f_{\rm in}}\right)^2} \dot{f}_{\rm in}
-   \frac{\sin{f_{\rm in}}\sqrt{1 - e_{\rm in}^2}  \left(2 + e_{\rm in} \cos{f_{\rm in}} \right)}
{\left(1 + e_{\rm in} \cos{f_{\rm in}} \right)^2}
\dot{e}_{\rm in}
.
\end{equation}

\noindent{}Finally, I use equation (\ref{dfdt}) to make the substitution for $\dot{f}_{\rm in}$, yielding

\begin{eqnarray}
\dot{\phi} &=& q_{_{\lambda_{\rm out}}}n_{\rm out} 
+ 
q_{_{\lambda_{\rm in}}}\frac{n_{\rm in} \left(1 + e_{\rm in} \cos{f_{\rm in}}\right)^2}{\left(1 - e_{\rm in}^2\right)^{3/2}}
\nonumber
\\
&-&
q_{_{\lambda_{\rm in}}}   
\frac{\sin{f_{\rm in}}\sqrt{1 - e_{\rm in}^2}  \left(2 + e_{\rm in} \cos{f_{\rm in}} \right)}
{\left(1 + e_{\rm in} \cos{f_{\rm in}} \right)^2}
\dot{e}_{\rm in}
\nonumber
\\
&+&
\left[
q_{_{\lambda_{\rm in}}}
\left(
1
- 
\frac{\left(1 - e_{\rm in}^2\right)^{3/2}}{\left(1 + e_{\rm in} \cos{f_{\rm in}}\right)^2}
\right)
+
q_{_{\varpi_{\rm in}}}
   \right]
\dot{\varpi}_{\rm in}
\nonumber
\\
&+&
\left[
q_{_{\lambda_{\rm in}}}
\left(
1
- 
\frac{\cos{i_{\rm in}}\left(1 - e_{\rm in}^2\right)^{3/2}}{\left(1 + e_{\rm in} \cos{f_{\rm in}}\right)^2}
\right)
+
q_{_{\varpi_{\rm in}}}
+
q_{_{\Omega_{\rm in}}}
   \right]
\dot{\Omega}_{\rm in}
.
\label{bigreseq}
\end{eqnarray}

\noindent{}Now one can use whichever set of equations for 
$\left\lbrace \dot{e}_{\rm in}, \dot{\omega}_{\rm in}, \dot{\Omega}_{\rm in} \right\rbrace$ from this 
paper which are appropriate to the system being studied.  Consequently, the time evolution of the 
resonant angle is expressed entirely in terms of orbital elements with no time derivatives on
the RHS.

Subsequently, regardless of whether one expands $\alpha$ about 0 to reduce the resulting equation, note there
is no need to expand separately about $e_{\rm in}$ nor $i_{\rm in}$ about 0 also.  Hence, one
can model resonant angles with high values of the eccentricity and inclination.

\section{Summary} \label{summ}

I have derived the equations of motion in the general restricted $N$-body problem as functions
of $\left(a,e,i,\Omega,\omega,f\right)$ only (equations \ref{firstdadt}-\ref{dfdt}).
I then expressed these relations with respect to an orbital plane that is fixed in space 
(equations \ref{dadtROT1}-\ref{dfdtROT1}), which represents a practical application in many contexts 
(e.g. a mass-hierarchic N-body problem, or the Solar System's ecliptic).
Modeling hyperbolic instead of elliptical orbits requires
only a change in the definition of $p_j$ (equation \ref{rjGEN}).  These equations, along with
their partially coplanar (equations \ref{dadtROT2}-\ref{dfdtROT2}) and fully coplanar 
(equations \ref{dadtPLANGENA}-\ref{dfdtplanar}) versions, are not subject to any averagings nor expansions nor
assumptions about small forces.  This formulation may facilitate the study of mean motion
resonances (equation \ref{bigreseq}).  Orbital elements measured with respect to the centres
of mass of a particular set of bodies require just a translation from the primary-centric
case (equations \ref{bary} and \ref{final}), although alternatively the explicit equations 
of motion may be used (equations \ref{dadtROT2sca}-\ref{dodtROT2scf}).

I applied some of the above equations to three-body systems, and presented singly- and doubly-averaged
expressions for nonplanar and planar configurations for the primary-centric (Section \ref{scsec}),
general barycentric (Section \ref{sfsec}), and circular circumbinary barycentric (Section \ref{circum}) cases.  
This procedure, can, for example, yield Lidov-Kozai terms 
to a desired order (equations \ref{Koze}-\ref{Kozw}).  The averaged equations reveal the dominant
drivers of orbital changes, and identify which elements remain stationary over long timescales
when measured with respect to the barycentre of the binary.  I also find exact solutions to leading
order (equations \ref{exact2} and \ref{genOm2avgcircum}-\ref{genom2avgcircum}) and to several orders 
(equation \ref{solveexact}) for the precession of the pericentre and node in circular circumbinary
systems with an eccentric external body.  A solution to leading order also exists
(equation \ref{doubavgp}) for coplanar wide binary systems.

\section*{Acknowledgments}

I thank the two referees for their assessments, which include the probing and helpful inquires of 
Michael Efroimsky.  I also thank Mark C. Wyatt for useful discussions.  This work benefited from 
support by the European Union through ERC grant number 320964.

\bibliographystyle{spbasic}

\begin{thebibliography}{}

\bibitem[Binney \& Tremaine(1987)]{bintre1987} 
Binney, J., \& Tremaine, S.: Galactic Dynamics.
Princeton University Press, Princeton (1987) 

\bibitem[Brouwer \& Clemence(1961)]{brocle1961}
Brouwer, D., Clemence, G.M.: Celestial Mechanics.
Academic Press, New York (1961)

\bibitem[Burns(1976)]{burns1976} Burns, J.A.\ 1976.
Elementary derivation of the perturbation equations of celestial mechanics.
Am. J. Phys. 44, 944-949.

\bibitem[Danby(1992)]{danby1992} Danby, J.~M.~A.:
Fundamentals of Celestial Mechanics. 
Willman-Bell, Richmond, VA (1992)

\bibitem[de la Fuente Marcos \& de la Fuente Marcos(2013)]{deletal2013} 
de la Fuente Marcos, C., \& de la Fuente Marcos, R. 2013.
The Chelyabinsk superbolide: a fragment of asteroid 2011 EO$_{40}$?
MNRAS, 436, L15-L19

\bibitem[Efroimsky \& Goldreich(2003)]{efrgol2003} Efroimsky, M., \&
Goldreich, P.\ 2003. 
Gauge symmetry of the N-body problem in the Hamilton-Jacobi approach.
Journal of Mathematical Physics, 44, 5958-5977.

\bibitem[Efroimsky \& Goldreich(2004)]{efrgol2004} Efroimsky, M., \&
Goldreich, P.\ 2004. 
Gauge freedom in the N-body problem of celestial mechanics.
A\&A, 415, 1187-1199.

\bibitem[Efroimsky(2005)]{efroimsky2005} Efroimsky, M.\ 2005.
Long-Term Evolution of orbits about A precessing oblate planet: 
1. The case of uniform precession.
Celestial Mechanics and Dynamical Astronomy, 91, 75-108.

\bibitem[G\'{o}mez et al.(2001)]{gometal2001} G\'{o}mez, G.,
Llibre, J., Mart\'{i}nez, R., Sim\'{o}, C.: 
Dynamics and Mission Design Near Libration Points.  World Scientific,
Singapore (2001)

\bibitem[Gurfil(2007)]{gurfil2007} Gurfil, P.\ 2007. 
Generalized solutions for relative spacecraft orbits under arbitrary perturbations.
Acta Astronautica, 60, 61-78.
 
\bibitem[Gurfil, Lainey \& Efroimsky(2007)]{guretal2007} 
Gurfil, P., Lainey, V., Efroimsky, M.\ 2007:
Long-term evolution of orbits about a precessing oblate planet: 3. A
semianalytical and a purely numerical approach.
Celestial Mechanics and Dynamical Astronomy, 99, 261-292.

\bibitem[Kozai(1962)]{kozai1962} Kozai, Y.\ 1962.  
Secular perturbations of asteroids with high inclination and eccentricity. Astronomical. J., 67, 591-598.

\bibitem[Lidov(1961)]{lidov1961} Lidov, M.L.\ 1961. Evolution of the planets artiﬁcial satellites orbits under effect of the outer bodies gravity perturbations. Artificial Satellites of the Earth (Moscow, USSR: Nauka Publishers) 8, 5–45.

\bibitem[Morbidelli(2002)]{morbidelli2002}
Morbidelli, A.: Modern Celestial Mechanics: Aspects of
Solar System Dynamics.  Taylor \& Francis, London (2002)

\bibitem[Murray \& Dermott(1999)]{murder1999} 
Murray, C.~D., \& Dermott, S.~F.: Solar System Dynamics.
Cambridge University Press, Cambridge (1999) 

\bibitem[Naoz et al.(2013)]{naoetal2013} Naoz, S., Farr, W.~M., 
Lithwick, Y., Rasio, F.~A., \& Teyssandier, J.\ 2013, 
Secular dynamics in hierarchical three-body systems.
MNRAS, 431, 2155-2171.

\bibitem[Roy(2005)]{roy2005} Roy, A.~E.: Orbital motion. 
Institute of Physics Publishing, Bristol (2005)

\bibitem[Schnittman(2010)]{schnittman2010} Schnittman, J.~D.\ 2010:
The Lagrange Equilibrium Points L$_4$ and L$_5$ in Black Hole Binary System
ApJ, 724, 39-48.

\bibitem[Shoemaker(1995)]{shoemaker1995} Shoemaker, E.~M.\ 1995:
Comet Shoemaker-Levy 9 at Jupiter. Geophysical Research Letters, 22, 1555.

\bibitem[Szebehely(1967)]{szebehely1967} Szebehely, V.: Theory of Orbits:
The Restricted Problem of Three Bodies.  Academic Press, New 
York (1967) 

\bibitem[Valtonen \& Karttunen(2006)]{valkar2006} Valtonen, M., 
\& Karttunen, H.: The Three-Body Problem.  Cambridge University Press,
Cambridge (2006)  

\bibitem[Veras \& Armitage(2007)]{verarm2007} Veras, D., \& Armitage, 
P.~J.\ 2007: Extrasolar planetary dynamics with a generalized planar Laplace-Lagrange secular theory, ApJ, 661, 1311-1322.

\bibitem[Veras \& Evans(2013)]{vereva2013} Veras, D., 
\& Evans, N.~W.\ 2013: Planetary orbital equations in externally-perturbed systems: 
position and velocity-dependent forces.  Celestial Mechanics and Dynamical 
Astronomy, 115, 123-141.

\bibitem[Wolszczan(1994)]{wolszczan1994} 
Wolszczan, A.\ 1994: Confirmation of Earth-Mass Planets Orbiting the Millisecond Pulsar PSR B1257+12 
Science, 264, 538-542.

\bibitem[Wolszczan \& Frail(1992)]{wolfra1992} 
Wolszczan, A., \& Frail, D.~A.\ 1992: A planetary system around the 
millisecond pulsar PSR1257 + 12, Nature, 355, 145-147.

\end{thebibliography}

\appendix

\section{Appendix:  Barycentric equations for an arbitrary number of bodies}

Here I derive the equations of motion in orbital elements for the secondary with respect to the barycentre
of 3 or more massive bodies.  This section is an extension of Section \ref{barythree}, which treats the
two-massive-body case.  Examples of physical situations in which the equations here may be applied include
a distant terrestrial planet orbiting in a circumbinary planetary system already containing a Hot Jupiter, 
or a free-floating distant comet or planet being gravitationally captured and kept by a trinary or 
quatenary stellar system.  The procedure entails achieving an equation similar in form to 
equations (\ref{eqmotion})-(\ref{deltajb}),
as in equation (\ref{bary2body}).  

Assume $\vec{r}$ is the vector from the primary to the (zero-mass) secondary, and $\vec{s}$ is the vector
from the secondary to the centre of mass of all the massive bodies.  I need to develop the key equation
of motion in terms of $\vec{s}$.  Assume the total number of 
bodies\footnote{The barycentre need not be with respect to all massive bodies of the system, 
usually just the massive bodies which are interior to the secondary.} in the system is 
$N \ge 3$, and for ease
of summation indexing, here let the primary be the first body (denoted with a subscript 1) and the secondary be the 
$N$th body. Further assume that the vector from the primary to the $u = 2...(N-1)$th body is $\vec{r}_{u}$, and the vector 
from the centre of mass of the system to the $j = 1...(N-1)$th body is denoted by $\vec{s}_{j}$.

The equation of motion for $\vec{r}$ is

\begin{equation}
\frac{d^2\vec{r}}{dt^2} 
= 
- \frac{G m_1 \vec{r}}{r^3}
+ 
\sum_{u=2}^{N-1}
\left[
G m_{u} \frac{\left(\vec{r}_{u} - \vec{r}\right)}
{\left|\vec{r}_{u} - \vec{r}\right|^3}
-
G m_{u} \frac{\vec{r}_{u}}{r_{u}^3}
\right]
.
\label{App1}
\end{equation}

\noindent{}I need to express equation (\ref{App1}) in terms of $\vec{s}$ and $\vec{r}_u$
for $u = 2...(N-1)$.  In order to make a substitution for $\vec{r}$ on the RHS, consider
the relation for the centre of mass of the system

\begin{equation}
\sum_{u=1}^{N-1} m_u \vec{s}_u = 0
.
\label{CM}
\end{equation}

\noindent{}Also, by simultaneously solving $(2N-4)$ vector triangle relations, along with equation
(\ref{CM}), I obtain

\begin{equation}
\vec{r} = 
\frac
{\sum_{u=2}^{N-1} m_u \vec{r}_u}
{\sum_{j=1}^{N-1} m_j}
-
\vec{s}
.
\label{App2}
\end{equation}

Equation (\ref{App2}) can be used on the RHS of equation (\ref{App1}).  To obtain an expression
for $d^2\vec{r}/dt^2$, I use both equation (\ref{App2}) and the following standard equations of
motion for $u = 2...(N-1)$ massive bodies

\begin{equation}
\frac{d^2\vec{r}_u}{dt^2} 
+
\frac{G\left(m_1 + m_u\right)\vec{r}_u}{r_{u}^3}
=
\sum_{{j = 2}\atop{j \ne u}}^{N-1}
G m_j
\left(
\frac{\left(\vec{r}_{j} - \vec{r}_u\right)}
{\left|\vec{r}_{j} - \vec{r}_u\right|^3}
-
\frac{\vec{r}_{j}}{r_{j}^3}
\right)
\end{equation}

\noindent{}to obtain

\begin{equation}
\frac{d^2\vec{r}}{dt^2}
=
-
\frac{d^2\vec{s}}{dt^2}
-
\sum_{u=2}^{N-1} G m_{u} \frac{\vec{r}_u}{r_{u}^3}
.
\label{App3}
\end{equation}

\noindent{}Finally, equations (\ref{CM}) and (\ref{App3}) give the desired form of equation (\ref{App1}) as

\begin{eqnarray}
\frac{d^2\vec{s}}{dt^2}
&=&
-\left(\sum_{u=1}^{N-1} Gm_u \right)
\frac{\vec{s}}{s^3}
+
Gm_1
\frac{
\frac
{\sum_{u=2}^{N-1} m_u \vec{r}_u}
{\sum_{j=1}^{N-1} m_j }
- 
\vec{s}
}
{
\left|
\frac
{\sum_{u=2}^{N-1} m_u \vec{r}_u}
{\sum_{j=1}^{N-1} m_j }
- 
\vec{s}
\right|^3
}
\nonumber
\\
&-&
\left[
\sum_{u=2}^{N-1} G m_u
\frac{
\frac
{ \left[ \left(\sum_{j=1}^{N-1} m_j \right) - m_u \right] \vec{r}_u
-
  \left[ \left(\sum_{j=2}^{N-1} m_j \vec{r}_j \right) - m_u \vec{r}_u  \right]
}
{\sum_{w = 1}^{N-1} m_w}
+
\vec{s}
}
{
\left|
\frac
{ \left[ \left(\sum_{j=1}^{N-1} m_j \right) - m_u \right] \vec{r}_u
-
  \left[ \left(\sum_{j=2}^{N-1} m_j \vec{r}_j \right) - m_u \vec{r}_u  \right]
}
{\sum_{w = 1}^{N-1} m_w}
+
\vec{s}
\right|^3
}
\right]
\nonumber
\\
&&
\nonumber
\\
&+& 
\left(\sum_{u=1}^{N-1} Gm_u \right)
\frac{\vec{s}}{s^3}
.
\end{eqnarray}

\noindent{}Let the orbital elements measured with respect to this centre of mass be denoted with a asterisk.
The equations of motion for an arbitrary variable $\beta^{\ast}$ are finally

\begin{eqnarray}
\frac{d\beta^{\ast}}{dt} &=&  
\frac{m_1}{m_{\rm t}} \left(\frac{d\beta}{dt}\right)_{{\rm t},B} 
\bigg|_{ \vec{r}_{\rm t} \rightarrow \left[\sum_{j = 1}^{N-1} m_j\right]^{-1} \left[ \sum_{u=2}^{N-1} m_u \vec{r}_u \right] }^{(a,e,i,\Omega,\omega,f) \rightarrow (a^{\ast},e^{\ast},i^{\ast},\Omega^{\ast},\omega^{\ast},f^{\ast})   }
\nonumber
\\
&&
\nonumber
\\
&+&
\left[
\sum_{u=2}^{N-1}
\left(\frac{d\beta}{dt}\right)_{u,B}
\bigg|_{ \vec{r}_{\rm u} \rightarrow -\left[\sum_{w = 1}^{N-1} m_w\right]^{-1} \left\lbrace      \left[ \left(\sum_{j=1}^{N-1} m_j \right) - m_u \right] \vec{r}_u - \left[ \left(\sum_{j=2}^{N-1} m_j \vec{r}_j \right) - m_u \vec{r}_u  \right]  \right \rbrace }^{(a,e,i,\Omega,\omega,f) \rightarrow (a^{\ast},e^{\ast},i^{\ast},\Omega^{\ast},\omega^{\ast},f^{\ast})   }
\right]
\nonumber
\\
&&
\nonumber
\\
&-&
\frac{\left(\sum_{j=1}^{N-1} m_j\right)}{m_{\rm t}}
\left(\frac{d\beta}{dt}\right)_{{\rm t},B}
\bigg|_{ \vec{r}_{\rm t} \rightarrow 0 }^{(a,e,i,\Omega,\omega,f) \rightarrow (a^{\ast},e^{\ast},i^{\ast},\Omega^{\ast},\omega^{\ast},f^{\ast})   }
.
\label{final}
\end{eqnarray}

\noindent{}In equation (\ref{final}), the tertiary-specific terms may be substituted for any body which is not the primary nor secondary.  Because the scalings in the first and second terms involve linear combinations of distances, the resulting expressions for the orbital element evolution will involve multiple distance ratios.

\end{document}